\documentclass[conference,usletter]{IEEEtran}
\IEEEoverridecommandlockouts
\usepackage{cite}
\usepackage[hyphens]{url}
\usepackage[colorlinks]{hyperref}
\usepackage{amsmath,amssymb,amsfonts}
\usepackage{algorithmic}
\usepackage{graphicx}
\usepackage{textcomp}
\usepackage{xcolor}
\usepackage{booktabs}
\usepackage{tabularx}
\usepackage{pgf}
\usepackage{import}
\usepackage{enumitem}
\usepackage{subfig}
\usepackage{siunitx}
\graphicspath{{./images/}}

\def\BibTeX{{\rm B\kern-.05em{\sc i\kern-.025em b}\kern-.08em
    T\kern-.1667em\lower.7ex\hbox{E}\kern-.125emX}}

\begin{document}

    \title{Pump, Dump, and then What? The Long-Term Impact of Cryptocurrency
Pump-and-Dump Schemes}

    \author{\IEEEauthorblockN{Joshua Clough}
	    \IEEEauthorblockN{Matthew Edwards}\\
        \IEEEauthorblockA{\textit{School of Computer Science} \\
        \textit{University of Bristol}\\
        Bristol, UK \\
        \{zs20903,matthew.john.edwards\}@bristol.ac.uk}
    }

 \maketitle{}

    \begin{abstract} 

The pump and dump scheme is a form of market manipulation attack in which
coordinated actors drive up the price of an asset in order to sell at a higher
price.  Due in part to a lack of enforcement, these schemes are widespread
within the cryptocurrency marketplace, but the negative impact of these events
on the coins they target is not yet fully understood.  Drawing upon a novel
dataset of pump events extracted from Telegram channels, an order of magnitude
larger than the nearest comparable dataset in the literature, we explore the
differing tactics of pumping channels and the long-term impact of pump and dump
schemes across 765 coins.  We find that, despite a short-term positive impact in
some cases, the long-term impact of pump and dump schemes on the targeted assets
is negative, amounting to an average 30\% relative drop in price a year after
the pump event.

    \end{abstract}

    \begin{IEEEkeywords}
    market manipulation, cryptocurrency, telegram, exchanges, fraud
    \end{IEEEkeywords}

    \section{Introduction}
    \label{sec:context}

\begin{figure}[!htbp]
    \centering
    \includegraphics[width=\linewidth]{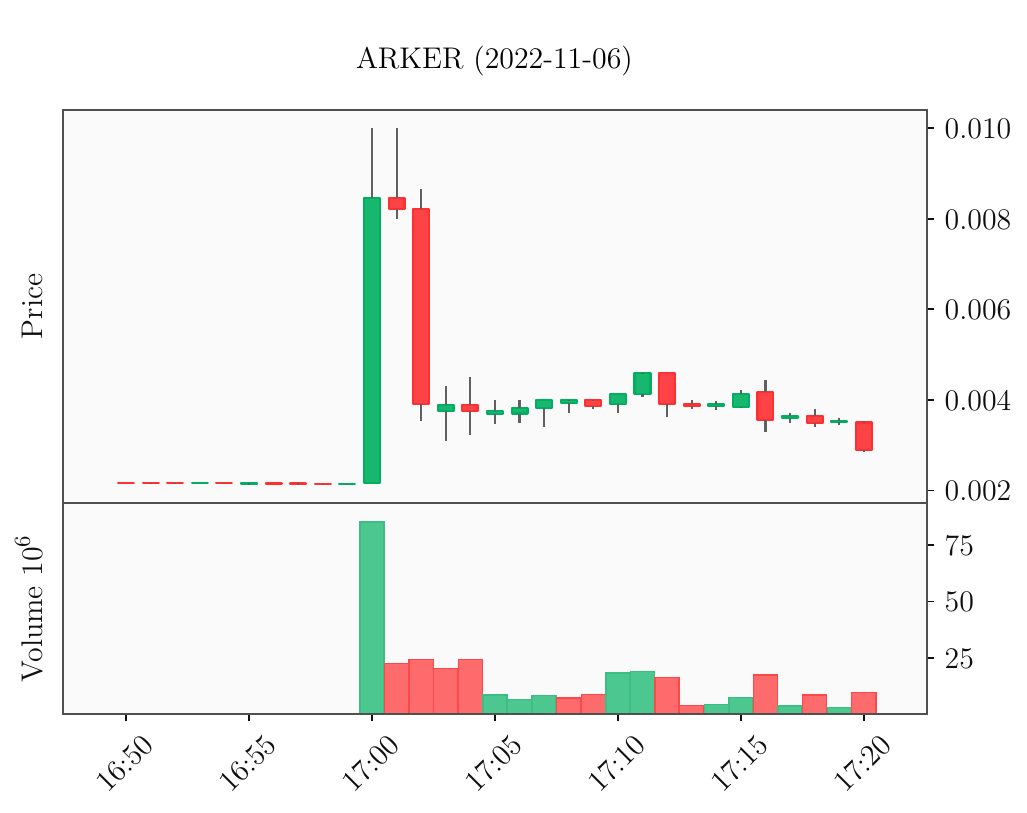}
    \caption{Example of a pump and dump event on the cryptocurrency ARKER.}
    \label{fig:arker_example}
\end{figure}

    Pump and dump schemes are a type of investment fraud where asset prices are artificially inflated by a group of market participants in order for them to sell the assets at a higher price.
Once the instigators sell off the asset and stop promoting it, the price falls significantly and any remaining investors in that particular asset end up losing money~\cite{SEC-pump-def-2020}.
Figure~\ref{fig:arker_example} shows an example of a cryptocurrency pump and dump event where the price rapidly rises as participants buy into the coin, peaks for around two minutes, then rapidly falls as some participants sell their holdings in the coin.

This type of scheme is not in itself a new phenomenon, with examples littering stock market history back to the South Sea Bubble of 1720.
However, the unregulated and decentralised nature of cryptocurrencies and the widespread adoption of encrypted messaging applications such as Telegram have made executing such schemes possible on a scale never seen before.
This near-industrial scale of such schemes means that they have had an increasingly significant impact on the cryptocurrency market as a whole, as such events can be organised on a daily basis and across multiple different exchanges.
The current state of regulation with respect to cryptocurrencies is sparse or non-existent~\cite{corbet-regulation-2021} and in effect this allows operators to get away with organising such events with no legal consequences or accountability, despite similar schemes in traditional stock markets being illegal with organisers actively prosecuted by the SEC~\cite{sec-market-charge-2022}.

The aims and contributions of this paper are:

\begin{itemize}
    \item \textbf{New and enlarged cryptocurrency pump event dataset.} This paper introduces and analyses a new enlarged dataset of pump events, which includes pump events collected directly from Telegram to build upon an existing dataset~\cite{morgia-doge-wall-street-2021}.
    Our new dataset expands this existing dataset from $1,111$ events to
$10,687$. The dataset is made publicly available for future
research\footnote{\url{https://gitlab.com/bristol-university-work/crypto-pump-and-dump}} along the code used to collect it, enabling future updates and expansions.
    \item \textbf{Pump strategy analysis.} A full breakdown and analysis of the dataset was performed with respect to the organising channel, exchange and market capitalisation, revealing two strategies taken by pump organisers, referred to as the \emph{quantity vs quality tradeoff}.
    A successful analysis of the performance of the pumps in the dataset further exhibits differences between these two groups.
    \item \textbf{Long-term impact.} Whilst it has been suggested by Li et al.\ \cite{li-pump-dump-schemes-2021} that a higher concentration of pump events occurring on a given exchange are detrimental to the price of cryptocurrencies in the long-term, no study has focused on concretely quantifying and analysing the pricing of pumped coins in the long-term fallout from such events.
    In our analysis of the pricing impact after the pumps within our dataset, it is revealed that prices of pumped coins fell by 30\% after 365 days relative to the wider market, indicating that such schemes have a strongly negative impact on cryptocurrency value.
\end{itemize}

    \section{Related Work}
    \label{sec:technical}

    \label{subsec:related-work}


Huang and Cheng~\cite{huang-stock-2015} investigate the impact of pump and dump events in the Taiwanese stock market, a regulated market, by analysing market data on manipulations prosecuted by authorities from 1990 to 2010.
They examine the cumulative abnormal returns (CAS) from day $-100$ to $+100$
relative to events and find that the peak CAS is $28\%$ but by day $+100$ the
CAS returns to $0\%$, suggesting that such events increase the volatility of returns during both pump and post-pump periods.
Analysing the effects on market prices they find that the temporary price increases prices by over $24\%$ but overall longer term price impacts are negligible.
They further show that there is also a large temporary price impact associated with such events and suggest that this means events have a damaging effect on price accuracy and market efficiency.

One of the first papers describing pump and dumps in the cryptocurrency sphere, Harmick et al.~\cite{hamrick-pump-dump-economics-2018} provide a description for pump and dump schemes and identify factors that affect the success of a pump.
They collected data over a 7 month period from Telegram and Discord and broadly
categorised the channels into 3 groups:  \textbf{Obvious pumps} that clearly
promote pump and dump schemes and provide countdown signals hours and days
before the occurrence of the pumps; \textbf{Target pumps.} that avoid directly
marketing themselves as a pump and dump channel but instead posted coin names
without any prior announcement and \textbf{copied pumps} that copied other
channels' posts, typically several hours after the original post.
They found that coins with lower trading volume (and therefore lower market capitalisation and liquidity) were more likely to produce a successful pump, and also established that the number of exchanges a coin is listed on correlates negatively with the success of a pump.


Klamps and Kleinberg~\cite{kamps-moon-2018} propose an unsupervised anomaly detection algorithm to help identify pump and dump events. They found that certain exchanges, specifically Binance and Bittrex, accounted for more pumps than the relative percentage of symbols explored on each whereas for Kraken, Kucoin and LBank the converse was true. On a coin-pair level, they found that most were targeted $0-3$ times but there were coins that were targeted up to 13 times, implying that pump and dump groups target specific coins multiple times.




Xu and Livshits~\cite{xu-anatomy-2019} provide an in depth analysis of the anatomy of pump and dump schemes and real-world case study of such an event. They note that participation levels in a group are at a fraction of the total membership. They go on to identify 412 pump events in 358 channels over an 8 month period on 4 exchanges.. After these pumps were matched to OHLCV market data using CryptoCompare, they develop a RF model and investment strategy for predicting pump events. They estimate that they only obtain half of the gain in value caused by the pump and find even with this caution, returns of $60\%$ can be achieved.

Li et al.~\cite{li-pump-dump-schemes-2021} investigated pump events using hand-collected data from Telegram across both CEXs (Binance, Bittrex and Yobit) and DEXs (PancakeSwap)\footnote{Not used in this project due to the difficulty in obtaining historical pricing data for DEXs.}, from an economic perspective.
They explain that these CEXs are not randomly chosen, and display common features such as having little if any ``know your customer'' requirements\footnote{Binance and Bittrex have since introduced such requirements.} and a large number of listed cryptocurrencies to use as targets.
After analysing market data around the collected events, they found that cryptocurrencies targeted are far more likely to have been pumped before and that effects on traded volumes disappear in one to two days, when viewed in the context on a week long window after events.
Further analysis on cryptocurrencies with a relatively high market capitalisation pumped on CEXs reveals that the effects of pump events are not restricted to only small coins and on average most pump events increase trading volumes and prices.
They conclude by analysing the effects of two opposing policy changes with respect to pump and dumps by Bittrex and Yobit.
In the case of Bittrex, which started banning accounts suspected of market manipulation in November 2017, the number of pump events sharply decreased.
Yobit, however, announced in October 2018 that it would randomly pump listed cryptocurrencies on its exchange, which generated a negative reaction from investors and reduced the overall prices and volumes of cryptocurrencies listed.
They suggest that these opposing effects mean that pump and dump events are damaging to the price and liquidity of cryptocurrencies.

Morgia et al.~\cite{morgia-doge-wall-street-2021} collected a publicly available dataset of pump events\footnote{\url{https://github.com/SystemsLab-Sapienza/pump-and-dump-dataset}, which forms part of the dataset used in this project.} and proposed a real-time detection model that represented a significant improvement on existing models with respect to speed and accuracy.
They also investigated ``crowd pumps''; pumps that result from actions by a group of market participants that are not directly organised.
They present pumps of GameStop and DogeCoin as examples of such events, and
compare ``crowd pumps'' to traditional pump and dump events, explaining that
there are three key differences between the two:  (1) crowd pumps aim to inflate
prices and keep them high, whereas traditional schemes quickly sell at inflated
prices for a profit; (2) the target of crowd pumps is known well in advance so
any uptick in its price can trigger the start of a large price increase and (3)
crowd pumps can last extended periods of time whereas traditional schemes
typically last minutes.



In other work, Victor and Hagermann~\cite{victor-quantification-detection-2019} analyse pumps on Binance over the period of a year and find that, on average, a pumped coin performs around $10\%$ better in the 100 days after a pump event compared to its peers.
They also use an XGBoost classifer, which uses tree boosting, to detect pump events and find $612$ pump like events across $172$ coins.
Corbet et al.~\cite{corbet-regulation-2021} find that cryptocurrencies do not fit into existing regulations.
This makes applying any existing market manipulation regulations to cryptocurrencies nigh on impossible.
Dhawan and Putniņš~\cite{dhawan-wolf-2021} postulate that no rational market participant would knowingly take part in a pump and dump scheme as they show them to be a negative-sum game.
They further explain that their evidence suggests participants treat such events as a game where the goal is to outsell others.

There are numerous other papers focussed on building machine learning models to detect pump and dump events.
Nilsen~\cite{nilsen-limelight-2019} used a LSTM network to create a real-time pump event detector with over $97\%$ accuracy, Tsuchiya~\cite{tsuchiya-profitability-2021} used Bayesian linear regression to classify pumps before they occur with a $75\%$ accuracy rate and Hu et al.~\cite{hu-sequence-2022} created a sequence-based neural network to identify pumps within Telegram channels.

Whilst this is clearly an active field, much previous work has focused on data collected during a specific period and for a specific exchange.
By collecting data going back multiple years and leveraging existing datasets, we produce a more comprehensive and extensive dataset that allows us to develop a novel long-term analysis, and which can be updated and adapted for future use.

    \section{Data Collection and Preparation}
    \label{sec:method_and_execution}


Our methodology is split into three distinct phases.

\begin{itemize}
    \item \textbf{Pump Event Collection (Section~\ref{subsec:pump-event-collection}).}
    Identifying Telegram channels organising pump and dump events and collecting information about such events from them.
    \item \textbf{Market Data Collection (Section~\ref{subsec:market-data-collection}).}
    Aggregating market OHLCV data for identified pump events.
    \item \textbf{Price and Data Analysis (Section~\ref{subsec:price-and-data-analysis}).}
    Using collected market data to analyse the impact of pump events.
\end{itemize}


    \subsection{Pump Event Collection}
\label{subsec:pump-event-collection}

There are few large-scale existing datasets on pump events, the lone exception
being the one produced by Morgia et al.~\cite{morgia-doge-wall-street-2021},
which spans multiple years (2017 to 2021), but covers only 1,111 pump events. 
We take this dataset as a basis, and update and extend it through our own process described below.

\subsubsection{Identifying Telegram Channels}
\label{subsubsec:identifying-telegram-channels}

We used PumpOlymp\footnote{\url{https://pumpolymp.com}.} to retrieve a list of
$800$ Telegram channels for investigation (as used by Xu and
Livshits~\cite{xu-anatomy-2019}). 
Further investigation on a subset of this list found that many of these channels
were inactive or provided ``signals'' rather than organising pump and dump
events. We filtered the list to find channel names that contained the word
``pump'', which yielded around $130$ channels for investigation.

\subsubsection{Finding and Collecting Pump Events}
\label{subsubsec:finding-pump-events}
\begin{figure}[!htbp]
    \centering
    \subfloat[Pump announcement using an image.] {
        \includegraphics[width=0.45\textwidth]{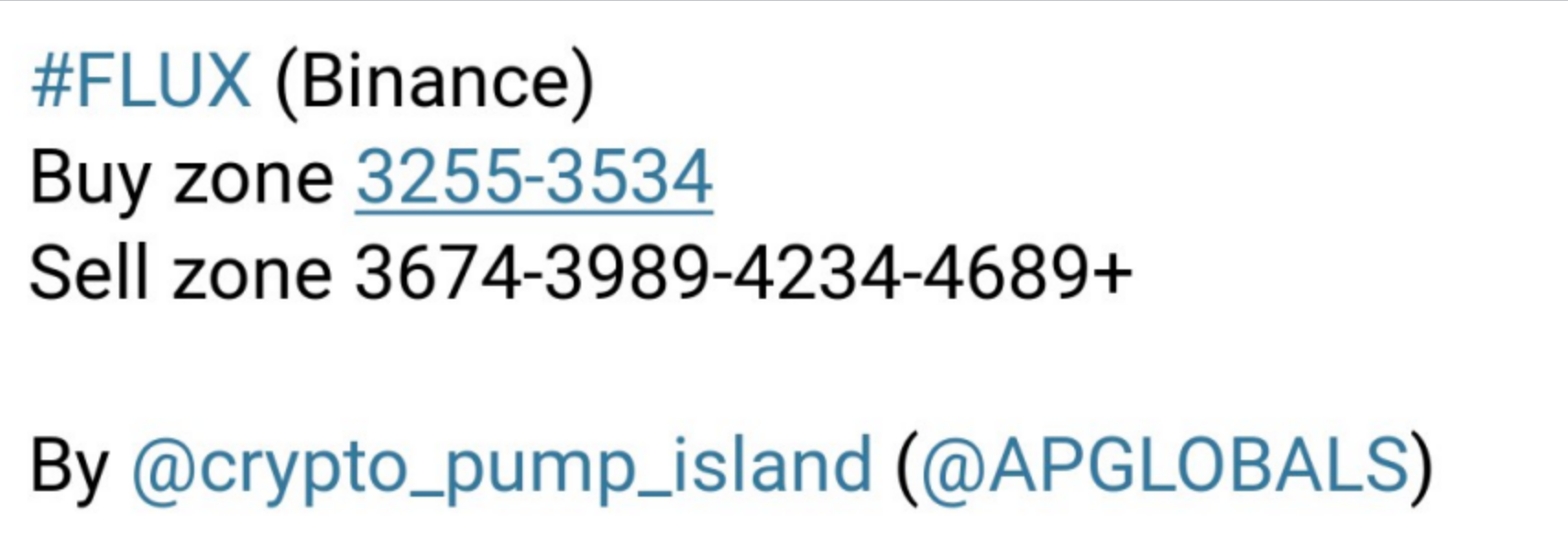}
        \label{fig:ocr_announcement}
    }
    \quad
    \subfloat[Pump announcement using text.] {
        \includegraphics[width=0.45\textwidth]{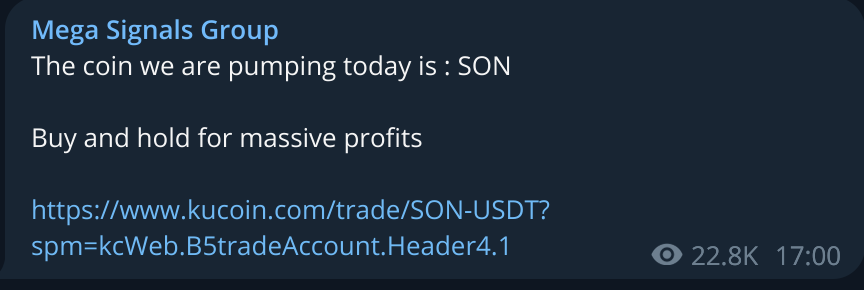}
        \label{fig:text_annoucement}
    }
    \caption{Different styles of pump announcements.}
    \label{fig:pump_announcements}
\end{figure}

Identifying pump events from channels was automated using the Telethon Python library.
The main challenge was the different formats used to announce coins being pumped.
As shown in Figure~\ref{fig:pump_announcements}, the announcements generally took one of two forms: text containing keywords (e.g., ``coin'' and ``pumping today'', as in Figure~\ref{fig:text_annoucement}), or text embedded in an image (as in Figure~\ref{fig:ocr_announcement}). We extracted text from images using pytesseract\footnote{\url{https://pypi.org/project/pytesseract}, powered by Google's Tesseract-OCR engine, \url{https://github.com/tesseract-ocr/tesseract}.} and iterated through images checking for predefined regex patterns (e.g., \verb|#|) matching identifiable patterns of announcement. 
These extracted coins were combined with the date and time of the message, also extracted via Telethon, to give a list of per-channel pump events.

Out of the $130$ channels investigated, $34$ of them had pump events that could be systematically identified using the methods described above.
There were multiple reasons for this large number of unusable channels, chief among them was channels either no longer existing or having broken links.
Issues also included channels organising pumps on DEXs, such as PancakeSwap, which do not have APIs giving access to historical OHLCV information. Some channels no longer organised pumps, instead providing general cryptocurrency investment ``advice'' in vague terms.

\subsubsection{Aggregating and Cleaning Pump Events}
\label{subsubsec:aggregating-pump-events-into-a-single-table}
The final stage of pump event collection was to clean and aggregate pump events collected from Telegram to ensure that individual pump events were attributable to their original channel and that market data for the coins used was retrievable.
This was split into three stages, detailed below.

\paragraph{Identifying Channels}
Identifying the individual channels and their respective pumps was done using two to four digit codes generated from the respective channel's names.
For example, the channel \verb|Hit Pump Angels| is represented by the code \verb|HPA|.
The full reference for this can be found in Table~\ref{tab:telegram_channels}, in Appendix~\ref{appx:data-sources}.
This was done to allow unique identification for pump events that were broadcast
across multiple channels and to ensure compatibility with the system used by
Morgia et al.'s dataset\footnote{See \url{https://github.com/SystemsLab-Sapienza/pump-and-dump-dataset/blob/master/groups.csv}.}.

\paragraph{Checking Out the Coins}
Once events were uniquely identified, they were merged into a main table containing $10,687$ pump events.
As there is a large amount of listing and delisting of cryptocurrencies on a weekly basis, the main table was then filtered to remove coins that were no longer listed on exchanges' APIs.
For example, in the week ending Sunday 19th March 2023, there were 27 new listings across Binance, Bitmart, Hotbit and Kucoin alone~\cite{crypto-listing-2023}.
This high turnover of listed coins made these checks essential.

Coin checks were done by comparing pumped coins and their respective exchanges against lists of cryptocurrencies available via CCXT, and by extension, the exchanges' APIs.
These checks revealed $459$ pump events with coins that were no longer listed, which were subsequently removed from the dataset.

\paragraph{The Curious Case of Yobit} Yobit is a popular exchange for executing pump events due to its lack of ``know your customer'' requirements\footnote{These are used by exchanges to verify the identity of customers.}~\cite{yobit-rules-2023}.
It has also previously performed its own random pumps on coins listed on its exchange~\cite{li-pump-dump-schemes-2021} which implies that the exchange wants to encourage pump and dump organisers to use their platform.
Unfortunately, Yobit's API only provides OHLCV data for the past 7 days which makes analysing events from several years ago impossible.
As such the $364$ pumps identified on Yobit were filtered out and excluded from the analysis.

Applying the above aggregation and cleaning steps gave $9,191$ pump events that were used for analysis in Section~\ref{sec:analysis}.

    \subsection{Market Data Collection}
\label{subsec:market-data-collection}
The next phase of the project was to retrieve market data for the pumps via CCXT\@.
This section explains the processes used to do this and reasoning behind their choice.

\subsubsection{Data Granularity}
\label{subsubsec:data-granularity}
Deciding on the amount of data to collect and its granularity was the key
decision for this stage.
CCXT, and the exchanges that it interfaces with, allow queries to OHLCV data at varying granularities, ranging from 1 minute intervals to 1 day and 1 month intervals\footnote{It is worth noting that many exchanges' APIs purport to only have 1 minute data for a specific period of time (typically 90 days) but no such restriction was found when collecting via CCXT.}.
In an ideal world, all data collected would be at 1 minute granularity to allow for the highest level of detail in the analysis.

There are some drawbacks to the approach outlined above, chief among them being the sheer amount of data required to capture multiple years worth of OHLCV history.
Furthermore, analysing volumes over periods longer than the minute (i.e., total volume over an entire day) requires additional calculation and overheads.
The solution to this was to retrieve different granularities at different time periods, with higher granularities over time periods closer to pump events.
Data was collected at 1 minute (1m) intervals for 1 day either side of a pump event, at 1 hour (1h) intervals for 1 week either side of the event, and at 1 day (1d) intervals from a coin's listing to the current day\footnote{For the purpose of this project this is 7th March 2023.}.
This gives both the benefits of being able to analyse trends over a long time period whilst also being able to see the immediate impacts of events at a high level of granularity.

\subsubsection{Back to the Start}
\label{subsubsec:back-to-the-start}
Collecting data at 1d intervals from a coin's listing date required this date to be found in an efficient way.
A basic solution to this would be to send a query spanning a time period from a year before cryptocurrencies became widely traded, such as 2005, and taking the first date data is available for as the listing date.
Due to limits on the number of values returned at once by CCXT (in order to
satisfy limits imposed by the underlying exchanges' APIs) it would be impossible
to implement this effectively without a significant query overhead.
To reduce the number of queries, we use a binary search\footnote{Adapted from
\url{https://gist.github.com/mr-easy/5185b1dcdd5f9f908ff196446f092e9b}.}, which
searches for listing dates by calculating and comparing a midpoint that is compared against the search value~\cite{miller-search-2018}.
By retrieving single OHLCV queries at midpoints, it is possible to tell whether a coin was listed before or after that date since if the query returns values, the coin was listed before the date used in the query and if it returns no values, then the coin was listed after the date of the query.
This method was applied to every entry in the pump dataset, giving every pumped coin a start date to collect data from.

\subsubsection{Turning the Tap On}
\label{subsubsec:turning-the-tap-on}
Once the listing dates of coins had been established, collecting the data at the granularities discussed in Section~\ref{subsubsec:data-granularity} was a relatively straightforward process.
Queries for the respective granularities were paginated to ensure that API request limits were not exceeded and multiple passes of the dataset were performed to ensure that no data was missed due to API rate limits.
Data was collected for each unique coin in the dataset of pumped coins, a total of $765$ coins, suggesting that coins were pumped on average around $12$ times across the dataset (further explored in Section~\ref{subsubsec:amount-of-pumps-per-coin}).

\subsubsection{A Cheeky Bit of BTC}
\label{subsubsec:a-cheeky-btc}
One issue with the price data was the variation in base cryptocurrencies for the pairs retrieved from the exchange.
For example, all the coins in the dataset pumped on Binance are paired with a \verb|BTC| base whereas all the coins pumped on Hotbit are paired with \verb|USDT|\footnote{A stablecoin that is designed to be valued at \$1.}.
The problem with this is that $1$ \verb|BTC| is worth around $26,900$ \verb|USDT|\footnote{As of the 28th March 2023, but this price obviously varies wildly.}, hence making any coin with a \verb|BTC| pairing seem $26,000$x lower in price when compared to a similarly valued coin with a \verb|USDT| pairing.
A simple fix to this would be to multiply all the \verb|BTC| paired prices by $26,000$ in order to achieve parity with \verb|USDT| paired coins.
Unfortunately, due to the massive fluctuations in the price between \verb|USDT|
and \verb|BTC|, this would introduce a large margin of error, particularly as the price of \verb|BTC| was as high as $60,000$ \verb|USDT| in 2021.

The solution was to collect OHLCV data for the \verb|BTC/USDT| pairing for the same timeframes and periods discussed in Section~\ref{subsubsec:data-granularity} for all coins with \verb|BTC| as their base pairing.
These prices can then be combined with the original \verb|BTC| base paired data to produce equivalent data in \verb|USDT|.
Combining the two sets of OHLCV was performed as follows

\begin{itemize}
    \item \textbf{Open} $-$ open price of the original \verb|BTC| based pair multiplied by open price of \verb|BTC/USDT| for the relevant date, timeframe and exchange.
    \item \textbf{High} $-$ high price of the original \verb|BTC| based pair multiplied by the typical price.
    \item \textbf{Low} $-$ low price of the original \verb|BTC| based pair multiplied by the typical price.
    \item \textbf{Volume} $-$ volume figures of the original \verb|BTC| based pair as this is quoted in the coin being pumped and is independent of any \verb|BTC/USDT| conversion.
\end{itemize}

\noindent Typical price is calculated as follows

\[ 
    \text{Typical Price} = \frac{\text{High} + \text{Low} + \text{Close}}{3} 
\]

\noindent which provides an average price across the respective timeframe~\cite{fidelity-typical-2020}.
This is used to reflect that high and low prices of the original \verb|BTC| based pairing and the \verb|BTC/USDT| pairing are unlikely to ever line up perfectly (i.e. occur at exactly the same time), especially for longer timeframes.
Therefore using an average of the \verb|BTC/USDT| conversion price somewhat mitigates against this whilst still capturing differences between the high and low prices in the original \verb|BTC| pairing data.

    \subsection{Price and Data Analysis}
\label{subsec:price-and-data-analysis}
The final phase of the project was to analyse the collected market data for
impacts and trends.  Whilst the results of this analysis are contained in
Section~\ref{sec:analysis}, we here explain certain measures used and their
justification.

\subsubsection{Market Capitalisation Data}
\label{subsubsec:market-capitalisation}
Market capitalisation is a measurement of an asset's value, calculated as the
multiplication of the current asset price by the amount of that asset in
circulation.
In the cryptocurrency sphere market capitalisation can be used as a relative measure of size and perceived levels of risk with respect to a cryptocurrency~\cite{coinbase-market-cap-2020}.
Cryptocurrencies with a high market capitalisation are perceived as less risky as they tend to have a history of growth and tend to be more liquid.
Market capitalisation also plays a role in the success of pump and dump events, with pumps targeting coins with a lower market capitalisation more likely to be successful~\cite{hamrick-pump-dump-economics-2018}.
As such, our analysis used market capitalisation as a way to compare coins targeted across exchanges.

However, each exchange has different prices and volumes for coins, hence only capturing a snapshot of the asset's true market capitalisation.
These price differentials are driven by the differences in trading volumes (and therefore liquidity), the fact that moving money across exchanges is inefficient and there is no accepted method for pricing such assets~\cite{cnbc-prices-2017}.

The solution was to use data from CoinMarketCap, a website that provides aggregated price and volume data for assets traded across multiple exchanges.
Prices for a given asset are calculated from a volume-weighted average of all the traded market pairs (e.g.\ BTC/USDT) for that asset which in turn are calculated by converting the price of a pair into USD using reference prices~\cite{cmc-price-2020}.
Similarly, volumes are calculated as the sum of an asset's volume across all
trading pairs where the volume of each trading pair is converted to USD using
reference prices\footnote{For more detail on the methodology used to calculate these, see \url{https://support.coinmarketcap.com/hc/en-us/articles/360043395912-Volume-Market-Pair-Cryptoasset-Exchange-Aggregate-}.}.
This price and volume data, along with aggregated market capitalisation data, was retrieved via CoinMarketCap's API\footnote{\url{https://coinmarketcap.com/api}.} on the 27th March 2023.

\subsubsection{Calculating Pre-Pump Data}
\label{subsubsec:calculating-pre-pump-data}
Values for pre-pump prices were calculated using the average closing price of the coin for the 7 days prior to the pump event.
The reason for this was to mitigate against the effects of any insiders such as channel admins or VIP members gaining knowledge of the coin about to be pumped before the specified start time and in effect pre-pumping the coin~\cite{hamrick-pump-dump-economics-2018,huang-stock-2015}.

\begin{figure}[!htbp]
    \centering
    \includegraphics[width=0.5\textwidth]{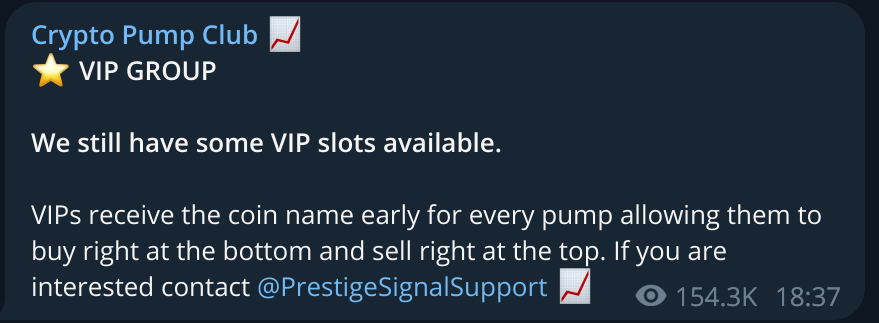}
    \caption{Example of VIP early access to pumps.}
    \label{fig:vip_early}
\end{figure}

Figure~\ref{fig:vip_early} shows a pump group advertising the VIP benefits of early access to coins being pumped, emphasising that prices of coins immediately prior to pumps have already been affected by such events.
7 days was chosen as the period to average over as it is long enough to smooth intraday market movements but short enough to not run into previous pump events.
Pre-pump volume data was similarly calculated as an average over the 7 days prior to a pump for consistency.

\subsubsection{Measuring Maximum Price Increase}
\label{subsubsec:measuring-maximum-price-increase} The maximum percentage price
increase $\Delta{P}$ used in our analysis is defined as

\[ \Delta{P} = \frac{P_{\max} - P_{\text{before}}}{P_{\text{before}}} * 100 \]

\noindent where

\begin{itemize}
    \item $P_{\max}$ is the maximum price in the 5 minutes after a pump announcement.
    \item $P_{\text{before}}$ is the pre-pump price defined in Section~\ref{subsubsec:calculating-pre-pump-data}.
\end{itemize}

\noindent We only use a 5 minute window to calculate this maximum increase
because price peaks of pump events are typically found within the first minutes
of pump event
announcements~\cite{victor-quantification-detection-2019,xu-anatomy-2019}.

\subsubsection{Measuring Volume}
\label{subsubsec:measuring-volume}




The volume within the pump window $\Delta{V}$ is calculated symmetrically to
$\Delta{P}$. Whilst the pre-pump and during pump volumes represent significantly
different timescales, the relative nature of this metric means that pumps across
coins with different liquidity characteristics can be directly compared.  The
volume moved during a pump event is taken as the amount of units of the coin
being pumped within the 5 minute window immediately after the announcement of a
pump event.  As discussed above, price peaks, and therefore the highest levels
of activity, are found within the first few minutes of such
announcements~\cite{victor-quantification-detection-2019,xu-anatomy-2019},
meaning that a 5 minute window is a reasonable period to use for this
calculation.

\subsubsection{Long-Term Timeframes}
\label{subsubsec:long-term-timeframes}
One of our key goals was to study impact of pump events over the long-term by
analysing price and volume data over multiple time periods after a pump event.
The timeframes used for analysis were 7, 14, 30, 60, 90, 180, 270 and 365 days,
to give a wide range of analysis that complements and extends existing research.
An important caveat to note is that there are pumps that occurred less than a
year before our market data was collected (e.g., pump events from January 2023
happened three months before our collection, hence there is only 3 months of
data to analyse rather than a year).  Another cause of censoring is that there
are coins for which subsequent pumps occur, at which point, for our analyses,
future data is considered to not be available for the original pump event, as
effects would otherwise be confounded by the second pump.

\subsubsection{Relative Price Impacts}
\label{subsubsec:relative-price-impacts}
In order to equally compare the price impacts across coins with massively different prices, a form of price indexing was used.
This set the pre-pump price of a pump event to 100, with subsequent relative prices calculated as follows

\[ R_{n} = \frac{P_{n}}{P_{pre}} * 100 \]

\noindent where

\begin{itemize}
    \item $R_{n}$ is the relative price at day n.
    \item $P_{n}$ is the absolute closing price at day n.
    \item $P_{pre}$ is the absolute pre-pump price for the pump, as calculated in Section~\ref{subsubsec:calculating-pre-pump-data}.
\end{itemize}

\noindent This means that subsequent absolute price rises above the pre-pump
price cause the relative price to rise above
100~\cite{economics-online-index-2020}.

\subsubsection{Real World Adjustments}
\label{subsubsec:real-world-adjustments}

We further compare the relative price indices of the pumps to the equivalent
relative indices of the top 10 cryptocurrencies by market
capitalisation\footnote{As of 22nd April 2023. Excludes stablecoins that are
meant to be pegged to a traditional currency.}.  These coins account for a large
proportion of the total market capitalisation of all cryptocurrencies, meaning
that their price movements represent the general sentiment of the market.  The
OHLCV data was retrieved at a daily granularity across the three most common
exchanges in the dataset; Binance, Kucoin and Hotbit.  The closing price was
averaged to give a combined price across all the exchanges for a given day.

These individual average prices were then converted to give relative prices, for
each of the top 10 coins, for the period after each pump event. Again, the
closing price for the day of pump was set to 100, with subsequent prices
calculated as detailed above.  The prices were combined through the use of a
weighted average, using the volumes of each coin as weights.  This gave average
relative market prices for the window after each pump, allowing a pumped coin's
price changes to be compared relative to the wider market.

\subsubsection{Stitching These Impacts Together}
\label{subsubsec:stiching-these-impacts-together}
Once every pump event had a relative price index for both the coin being pumped and the top 10 cryptocurrencies, the next stage was to combine these to allow the impacts to be measured.
This was calculated as follows

\[ I_{n} = (R_{n} - M_{n}) * 100 \]

\noindent where

\begin{itemize}
    \item $I_{n}$ is the resulting relative price adjusted for market price movements at day n.
    \item $R_{n}$ is the relative price of the pumped coins at day n.
    \item $M_{n}$ is the relative market prices of pumped coins at day n.
    \item $100$ is a normalising factor to move the resulting price difference back to 100.
\end{itemize}

The result of this was relative prices adjusted for market movements for each
pumped coin, allowing for a comparison of the price impacts independent of
general market movements.  For example, if a pumped coin is not affected at all
and follows general market movements, its adjusted price would be close to or
equal to 100 for the entire period.  Adjusted relative prices were averaged
across the entire dataset to give an average price change relative to the
pre-pump price and general market movements.  In order to do this, three
different averages were used in order to capture the effect of outliers and view
the overall data from different angles: the mean adjusted relative price for a
given day, the median, and the mean of all values within the interquartile
range. 





\subsubsection{Quantifying the Effect of Subsequent Pumps}
\label{subsubsec:quantifying-subsequent-pumps}
The final part of investigating the long term impacts was exploring the price differentials between coins with varying numbers of pumps, which was achieved via an extension of the method discussed above.
The first step required to achieve this was to put the pumps into a number of
predetermined bins, where each bin contains a similar number of pumped coins.
These are based on the number of times coins are pumped in the dataset.  Since
we found two groups in the dataset with significant differences in re-pumping
behaviour, there are two separate sets of bins.  These two groups are
distinguished by the groups organising them and are labelled as CPI organised
and non-CPI organised pumps.  A full breakdown of the differences between these
groups can be found in Section~\ref{subsubsec:quality-over-quantity}.

For CPI organised pumps, four bins were chosen, \verb|1-10| pumps, \verb|11-18|
pumps, \verb|19-30| pumps and \verb|31+| pumps, with each bin containing a
similar number of coins.  For non-CPI organised pumps, for which re-pumping was
rarer, two bins were chosen, \verb|1| pump and \verb|2+| pumps.  

Following binning, an average adjusted relative price was calculated for the 365
days after pump events, similar to
Section~\ref{subsubsec:stiching-these-impacts-together}.  One difference,
however, was that data for the entire 365 day time period was used for every
pump, instead of stopping if there was a subsequent pump.  As with above, the
mean, median and IQR mean were used as comparative averages.  We also compared
the performance of a coin's first pump to subsequent ones.  Again this analysis
was split for CPI and non-CPI pump events in order to better differentiate
between the two different organiser behaviours.  For both groups, we compare
performance ranging from the 1st to the 4th pump of a coin.




    \section{Analysis}
    \label{sec:analysis}

    \subsection{Dataset Breakdown}
\label{subsec:dataset-analysis-and-breakdown}
The first section of analysis breaks down the pump event dataset and highlights
some initial features and trends contained within it, including a natural split
in the data which is used as a segregating feature in future sections.

\subsubsection{Distribution of Pumps}
\label{subsubsec:distribution-of-pumps}

\begin{figure*}[!h]
    \centering
    \includegraphics[width=0.8\textwidth]{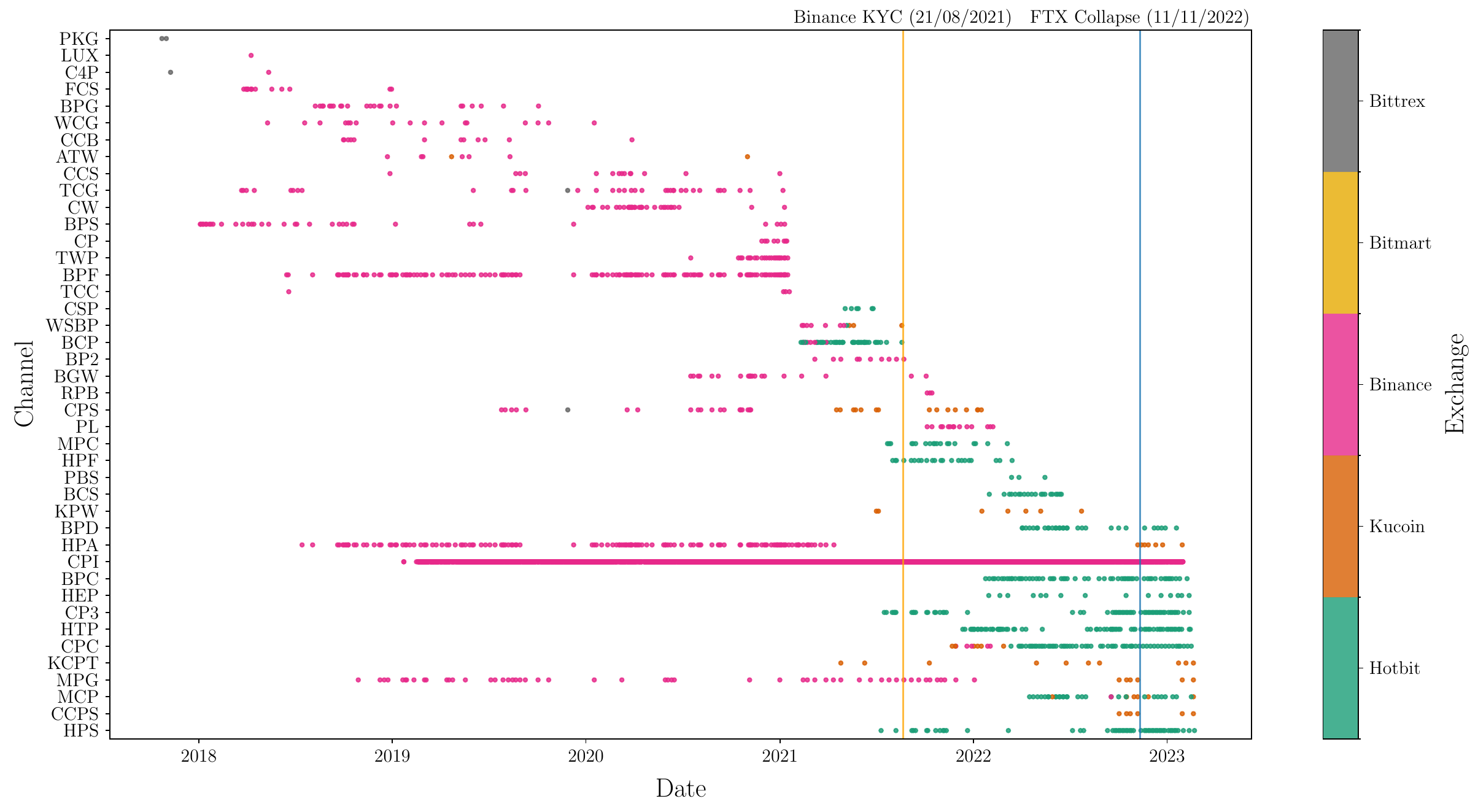}
    \caption{Distribution of pumps categorised by channel and exchange.}
    \label{fig:pump_distribution}
\end{figure*}

Figure~\ref{fig:pump_distribution} visualises the spread of pumps across the time period covered by the dataset\footnote{Inspired by Morgia et al.~\cite{morgia-doge-wall-street-2021}, see page 7, Figure 3.}.
The date of FTX's collapse is shown by the blue vertical line\footnote{The date FTX filed for bankruptcy: the 11th of November, 2022.}.
The plot indicates that there has been a shift in the exchange of choice for pump organisers from Binance to Hotbit and Kucoin, potentially caused by the introduction of mandatory KYC checks by Binance in 2021~\cite{binance-kyc-2021}, marked by the orange line.
This change in policy meant that every user that wanted to trade or deposit funds had to pass some form of KYC, a feature that is not attractive to pump and dump groups~\cite{li-pump-dump-schemes-2021}.
On the other hand, Kucoin allows users to withdraw up to 1 \verb|BTC|\footnote{Equivalent to around 26,900 USDT on the 28th March 2023.} a day and perform unlimited amounts of trades without any KYC~\cite{kucoin-kyc-2022} and Hotbit only requires KYC if a user triggers a ``higher risk control system''~\cite{hotbit-kyc-2023}.
This lack of KYC requirements makes it easier for participants to take part in pump events on the respective exchanges, making them more attractive to groups organising pump events.

\subsubsection{Quality over Quantity?}
\label{subsubsec:quality-over-quantity}

\begin{table}[!htbp]
    \centering
    \caption{Number of pumps per channel in the dataset.}
    \label{tab:num_pumps_per_channel}
    \begin{tabular}{llrrr}
        \toprule
        \textbf{Code} & \textbf{Channel Name} & \textbf{Pumps} & \textbf{} & \textbf{per Coin} \\
        \midrule
        CPI & Crypto Pump Island & 7920 & 86.17 & 22.06 \\
        HPA & Hit Pump Angels & 145 & 1.58 & 2.20 \\
        BPF & Binance Pump Family & 139 & 1.51 & 2.21 \\
        SP  & Softex Pump & 71 & 0.77 & 1.06 \\
        CPC & Crypto Pump Club & 70 & 0.76 & 1.15 \\
        HTP & Hotbit Trading Pump & 70 & 0.76 & 1.06 \\
        Others & & 776 & 8.45 & 1.16 \\
        \midrule
        \textbf{Total}   & & \textbf{9191} & & \\
        \bottomrule
    \end{tabular}
\end{table}

There is an outlier to this behaviour, a channel named Crypto Pump Island (CPI)\footnote{A full reference for channels and their respective channel codes can be found in Table~\ref{tab:telegram_channels}, in Appendix~\ref{appx:data-sources}} which has continued to organise pumps on Binance at a very high intensity.
Table~\ref{tab:num_pumps_per_channel} shows that CPI is responsible for over $86\%$ of the pumps in the dataset and has the highest number of pumps per coin by a factor of 10, meaning that each unique coin has been pumped by that channel on average $22$ times across the time period of the dataset.
This suggests that CPI is recycling coins when organising pumps, often multiple times a day, and is trying to achieve a high quantity of pumps rather than performance quality.

\subsubsection{Pumps per Coin}
\label{subsubsec:amount-of-pumps-per-coin}

\begin{table}[!htbp]
    \centering
    \caption{Number of pumps per coin in the datset.}
    \label{tab:num_pumps_per_coin}
    \begin{tabular}{lrr}
        \toprule
        \textbf{Number of Pumps} &  \textbf{Coins} & \textbf{Percentage} \\
        \midrule
        1--10  &   480 &       64.26 \\
        11--20 &   104 &       13.92 \\
        21--30 &    60 &        8.03 \\
        31--40 &    42 &        5.62 \\
        41--50 &    21 &        2.81 \\
        51+   &    40 &        5.35 \\
        \bottomrule
    \end{tabular}
\end{table}

There are only $765$ unique coins in the dataset of $9,000$ events, meaning coins were pumped multiple times.
Table~\ref{tab:num_pumps_per_coin} shows the distribution of the number of pumps per coin across the dataset.
Whilst most coins were targeted less than 10 times, there is a somewhat significant number of coins targeted over 30 times, likely for pumps organised by CPI\@.
The most targeted coin has $98$ pumps across a 4 year period.
This means that there are coins that are frequently targeted by operators of pump and dump schemes, suggesting they have a track record of success and they have features that make them attractive to such schemes.

\subsubsection{Market Capitalisation}
\label{subsubsec:market-capitlisation}
\begin{figure}[!htbp]
    \centering
    \includegraphics[width=0.5\textwidth]{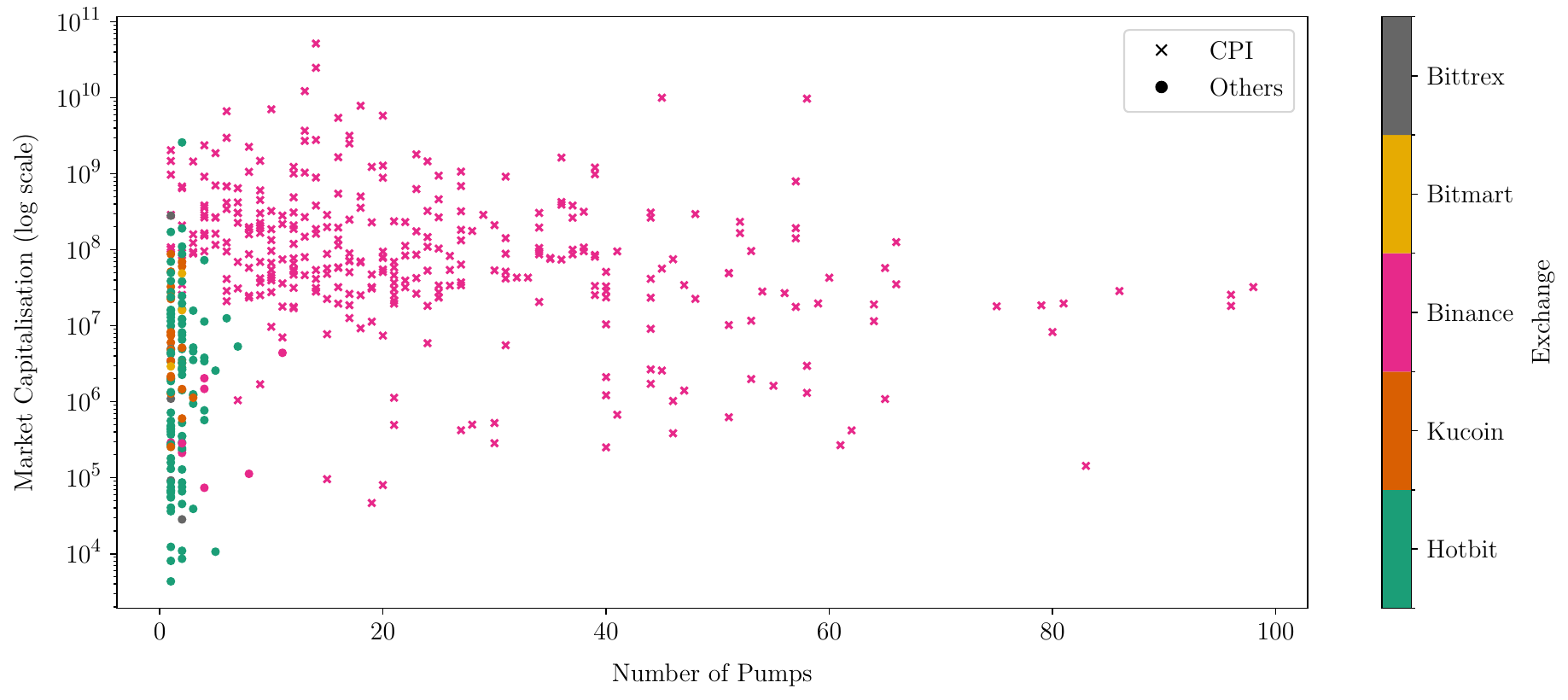}
    \caption{Number of pumps against the coins' respective market capitalisation.}
    \label{fig:market_cap}
\end{figure}

One such attractive feature could be a low market capitalisation, which has been linked to a pump's success~\cite{hamrick-pump-dump-economics-2018}.
Figure~\ref{fig:market_cap} shows the number of pumps per coin plotted against the coin's respective market capitalisation which are further segregated by the exchange in which the highest number of pumps took place on. It also highlights whether the coin was pumped predominantly on CPI (\verb|x| markers) or other channels (\verb|o| markers).

Coins that are pumped more often in general have a higher market capitalisation, occur on Binance and are typically organised by CPI\@.
As discussed in Section~\ref{subsubsec:quality-over-quantity}, CPI runs multiple
pumps a day which implies a preference for more liquid coins, as it is easier for participants to execute trades in the market~\cite{cmc-liquid-2021} meaning buying into a pump is easier.
Coins with higher liquidity tend to have a higher market capitalisation than less liquid ones which explains the behaviour in Figure~\ref{fig:market_cap}.
However, this higher liquidity and market capitalisation comes at the cost of lower returns on pumps~\cite{hamrick-pump-dump-economics-2018}, again highlighting the quality versus quantity tradeoff.

    \subsection{Pump Performance}
\label{subsec:pump-performance}
The second section of analysis investigates the relationships between pre-pump characteristics and the immediate performance of pumps.
It further highlights the quality vs quantity distinction defined in Section~\ref{subsec:dataset-analysis-and-breakdown} and explores the total amount of value moved .

\subsubsection{Pre-Pump Prices and Volumes}
\label{subsubsec:pre-pump-prices-and-volumes}
\begin{figure}[!htbp]
    \centering
    \includegraphics[width=0.5\textwidth]{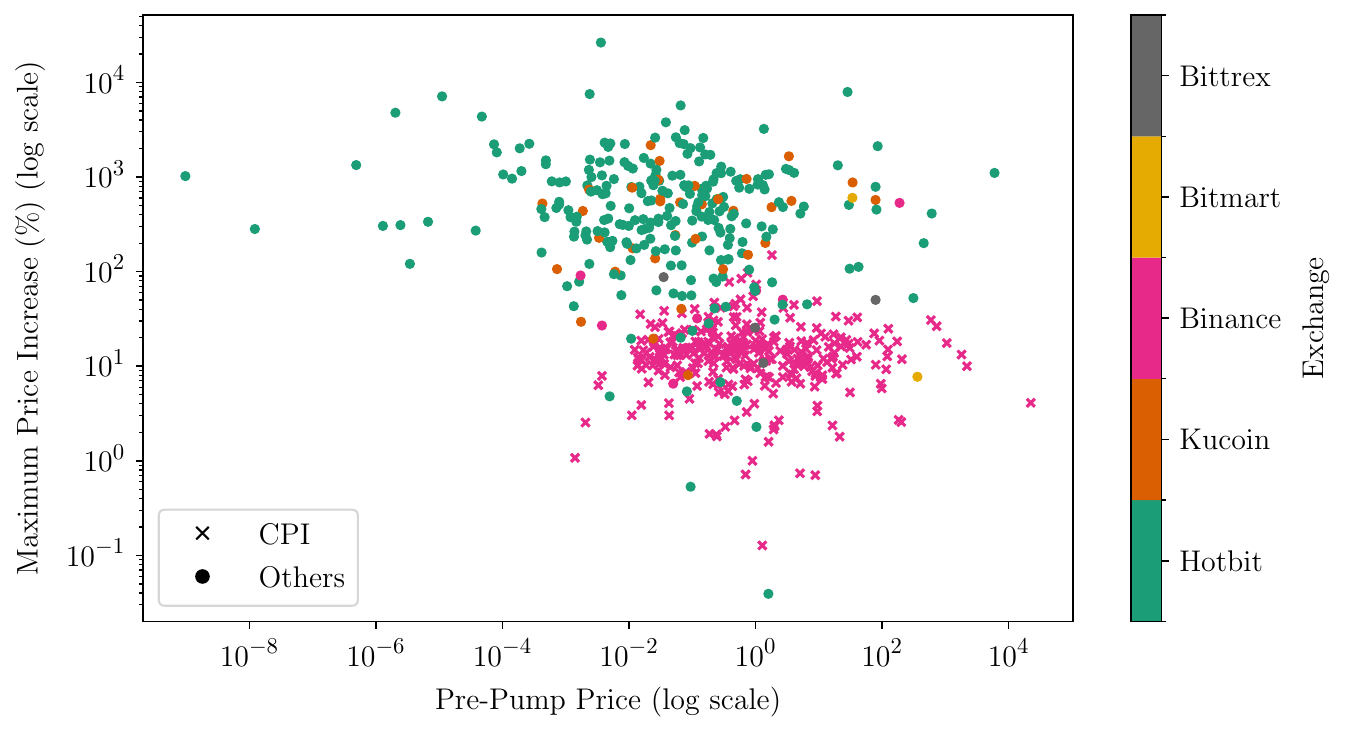}
    \caption{Pre-price of coins against the average percentage price increase caused by pump events.}
    \label{fig:increase_pre_price}
\end{figure}

Figure~\ref{fig:increase_pre_price} plots the pre-pump price against the average maximum price increase of pumps for each unique coin.
Once again there is clear separation between pumps organised by CPI and those organised by others.
The maximum price increase, on average, is much lower ($15.02\%$ vs $790.54\%$) for pumps organised by CPI and therefore for those taking place on Binance.
Conversely the pre-pump price, on average, is higher ($92.19$ USDT vs $41.01$ USDT) for pumps organised by CPI\@.

The negative trade-off between pre-price and increase is in part explained by the fact that a higher pre-price means a coin's price has to increase more in absolute terms, in order to achieve the same percentage increase as a coin with a lower pre-price.
Also the effects of choosing coins with higher market capitalisation, and therefore higher liquidity, mean the proportion of market participants attempting to push the price up is lower.
This means that it is harder to generate the volume required to massively increase prices.



\subsubsection{Volume Moved}
\label{subsubsec:volume-moved}
\begin{figure}[!htbp]
    \centering
    \includegraphics[width=0.5\textwidth]{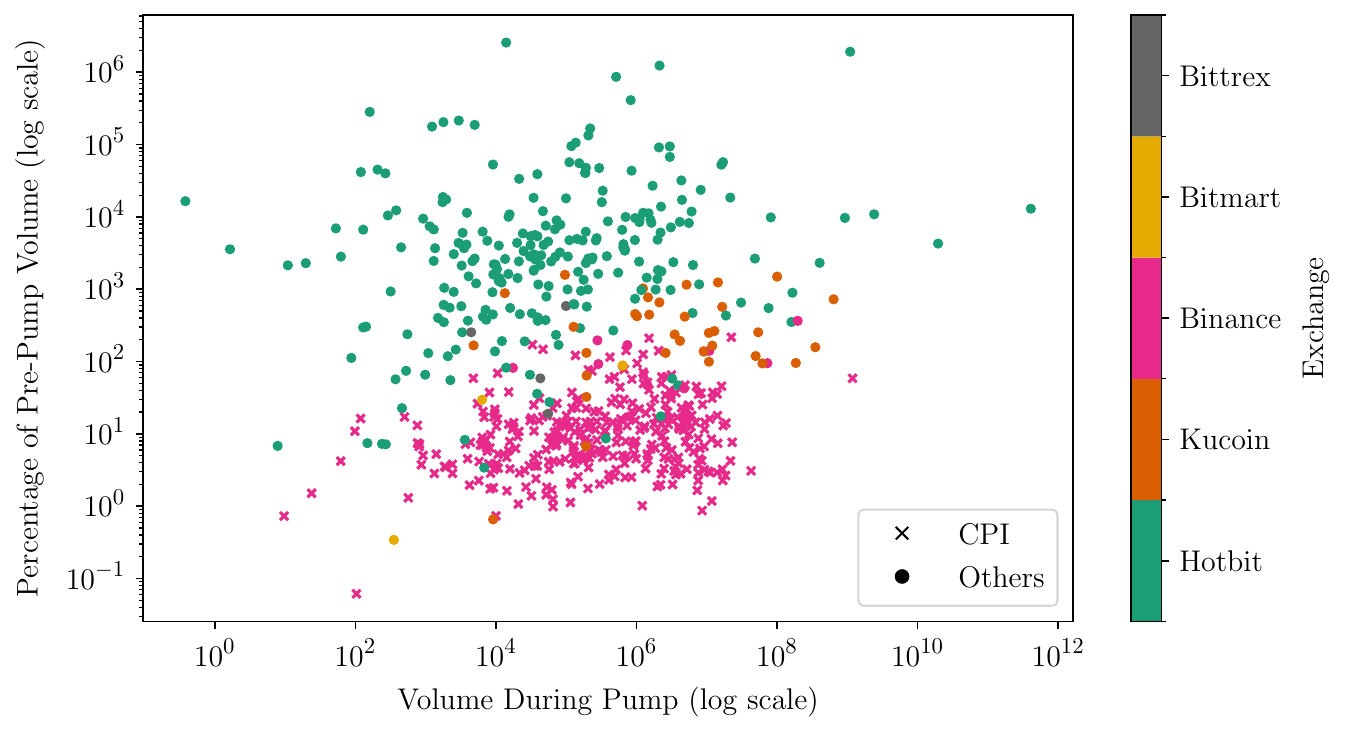}
    \caption{Volume moved during a pump event against the proportion of this relative to the pre-pump volume.}
    \label{fig:volume_diff}
\end{figure}

Figure~\ref{fig:volume_diff} shows the average volume moved in the first 5 minutes after a pump event announcement, compared to the proportion of the pre-pump volume this represents.
The amount of volume moved during a pump appears relatively similar for CPI and non-CPI pumps, although there is more variation and a higher standard deviation for non-CPI pumps ($\num{2.33e10}$ vs $\num{6.38e7}$).
It also shows that the average proportion of the pre-pump volume this represents is significantly higher for non-CPI organised pumps ($38703\%$ vs $17.38\%$), although again the non-CPI pumps have a much higher standard deviation for this metric ($25010$ vs $27.67$).
Looking at the median values ($1613\%$ for non-CPI, $8.25\%$ for CPI) it is evident that pumps organised by CPI have a significantly lower impact on the increase in volumes.
The choice of coins with higher trading pre-pump volumes and prices by CPI, as discussed earlier, means that more volume needs to be moved in order to reach the same values achieved by coins with very low pre-pump prices and volumes favoured by non-CPI organised pumps.

\subsubsection{Total Value}
\label{subsubsec:value-added}
Up to this point this section has been exploring the relative impacts of pump events.
Whilst this allows pump events to be compared on a like for like basis, it does not capture the total changes in value caused by pump events.
For example, the total value of trades for 10 units on a coin worth 1 USDT is 10 USDT but for a coin worth 10 USDT the total trading value is 100 USDT for the same 10 units traded.
To estimate the total value of a pump, we use the average maximum price achieved in the 5 minutes after a pump announcement.
Whilst this is obviously not the price across the entire 5 minute period, it produces the best possible price, and therefore the best possible total value of trades across the 5 minute period, which can be compared across all pump events.
The pre-pump total trading value is calculated as the pre-pump price multiplied by the pre-pump volume.

\begin{figure}[!htbp]
    \centering
    \includegraphics[width=0.5\textwidth]{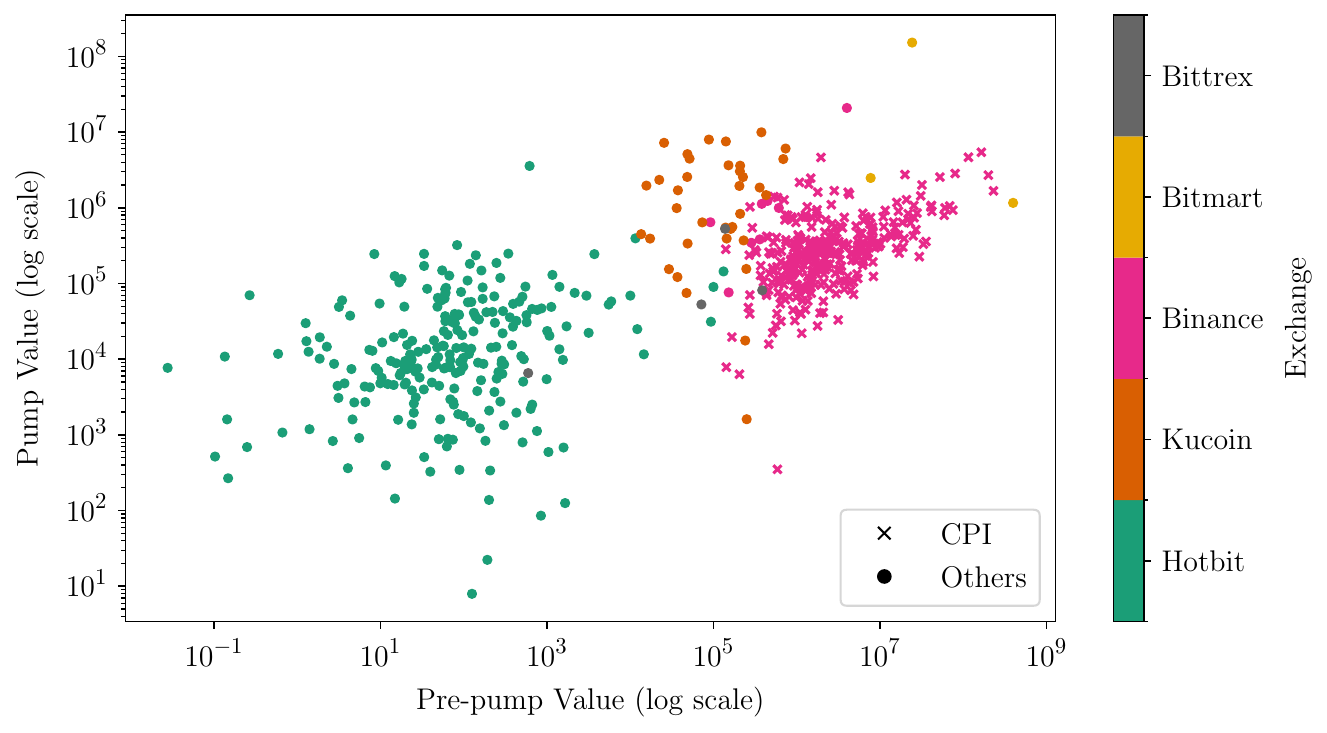}
    \caption{Total value of trades during pump events against to the average daily value of trades pre-pump.}
    \label{fig:pump_value}
\end{figure}

Figure~\ref{fig:pump_value} shows the total value of trades in the 5 minutes after a pump announcement compared to the average daily value of trades in the pre-pump period.
This indicates that the average daily value of trades is higher for pumps organised by CPI and lower for those not, which is not surprising given the higher pre-pump prices and volumes for coins pumped by CPI discussed earlier.
The value of trades in the 5 minutes after a pump announcement is also generally higher for pumps organised by CPI, particularly when compared to those organised by Hotbit.
Interestingly, pumps on Kucoin have the highest average value of trades in the 5 minutes after a pump announcement.
This is likely due to the Big Pump Signal group, one of the largest groups with some of the biggest price increases~\cite{morgia-doge-wall-street-2021}, using Kucoin as its primary exchange.
The large number of participants in pumps organised by this group means a higher volume of a coin is traded which in turn allows the price to be pushed up higher, which increases the total value of trades after a pump announcement.

    \subsection{Long-Term Impacts}
\label{subsec:long-term-impacts}
The third section of analysis focuses on the long-term impact of pump events on cryptocurrencies both overall and separated by pump organiser.
It also investigates the notion of a relationship between the number of pumps and long-term price performance.

\subsubsection{Overall Relative Impact}
\label{subsubsec:relative-impact}

\begin{figure}[!htbp]
    \begin{center}
            \includegraphics[width=0.8\linewidth]{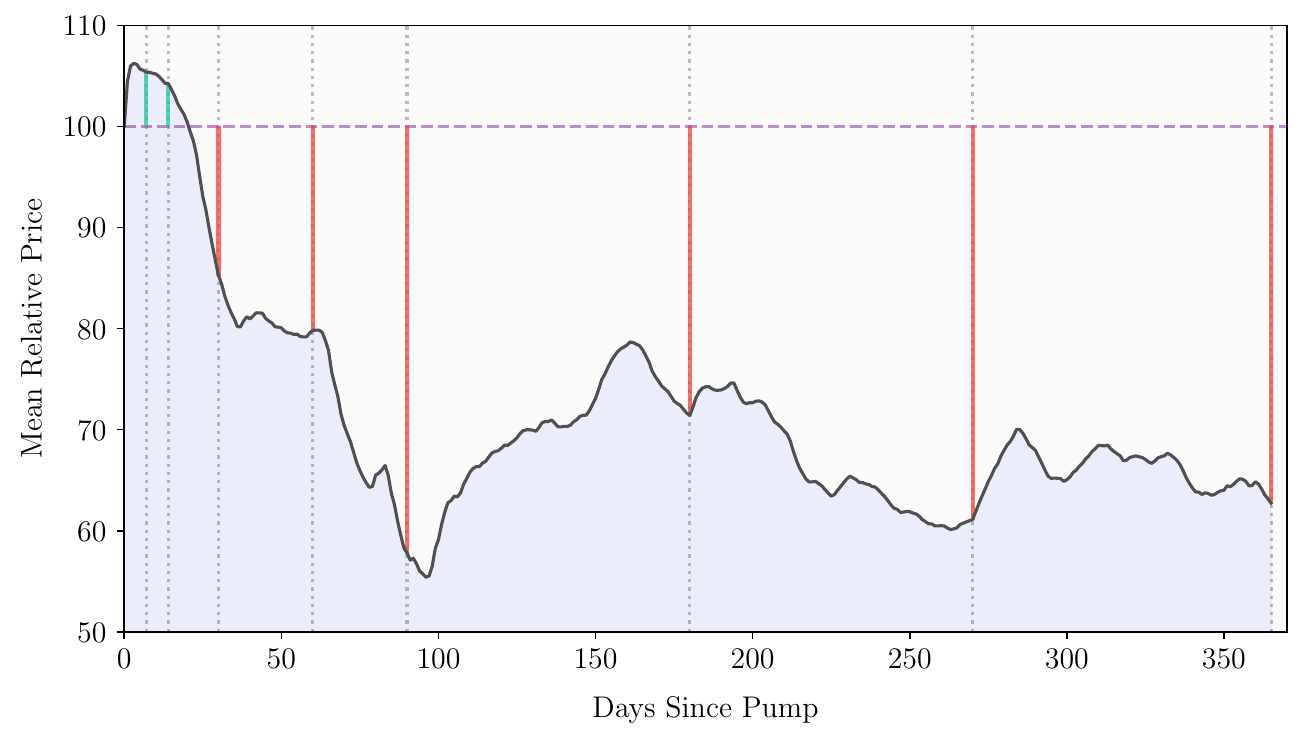}
        \caption{Mean relative prices of pumped coins in the year after a pump, adjusted for general market movements.}
        \label{fig:diff_all}
    \end{center}
\end{figure}

Figure~\ref{fig:diff_all} shows the mean prices of all pumped coins, relative to the pre-pump price, for the period spanning the year after a pump event.
This is compared to the relative prices of the top 10 coins by market capitalisation, referred to as the market prices from this point onwards.
This data is derived using the methodology outlined in Section~\ref{subsubsec:stiching-these-impacts-together} and it is worth noting that all relative prices for the coins are given with respect to the market prices.
It also highlights the percentage difference between the pumped coins price and the market price at different timeframes throughout the year, shown via the vertical lines.
Red lines indicate a percentage decrease relative to the market price and green lines indicate a percentage increase relative to the market price.

\begin{table}[!htbp]
    \centering
    \caption{Summary of the long-term price impact relative to a normalised pre-pump price of 100.}
    \label{tab:diff_all}
    \addtolength{\tabcolsep}{-0.4em}
    \begin{tabular}{lrrrrrrrr}
        \toprule
        Day & \textbf{7} & \textbf{14} & \textbf{30} & \textbf{60} & \textbf{90} & \textbf{180} & \textbf{270} & \textbf{365} \\
        \midrule
        Mean & 104.04 & 100.01 & 79.73 & 82.15 & 62.87 & 69.32 & 60.83 & 59.23 \\
        Median & 102.86 & 100.35 & 93.33 & 87.81 & 81.52 & 73.05 & 74.03 & 72.63 \\
        IQR Mean & 103.16 & 100.44 & 94.96 & 89.36 & 82.86 & 77.33 & 78.32 & 79.84 \\
        \midrule
        Average & 103.35 & 100.27 & 89.34 & 86.44 & 75.75 & 73.23 & 71.06 & 70.57 \\
        \bottomrule
    \end{tabular}
\end{table}

In the short term Figure~\ref{fig:diff_all} shows that pumps have a positive pricing impact relative to the market and pre-pump pricing.
This positive effect, however, is short-lived and the price is on average 11\% lower than the pre-pump value, relative to market prices, 30 days after the event.
Table~\ref{tab:diff_all} further shows the relative prices for each average at the timeframes highlighted in the figure.
It further emphasises that there is a small positive increase in price performance relative to market prices in the first week after a pump, on average around 3\%.
Expanding the timeframe out gives a decrease of around 15\% at 60 days and 30\% at 365 days, implying a steady decline in the pumped coin's value relative to market prices.

\subsubsection{Impact by Organiser}
\label{subsubsec:organiser-impact}

\begin{figure}[!htbp]
    \begin{center}
            \includegraphics[width=0.8\linewidth]{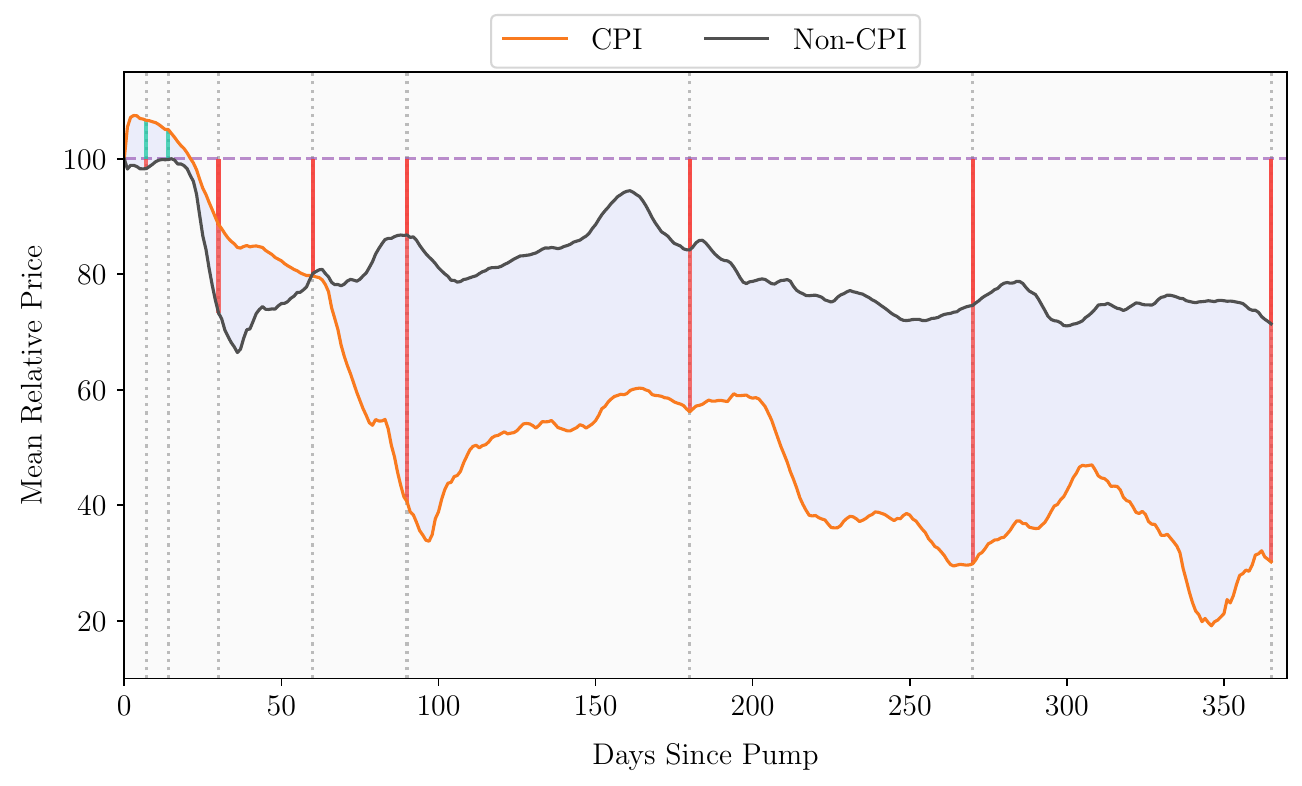}
        \caption{Mean relative prices of pumped coins in the year after a pump, adjusted for general market movements and separated by organiser.}
        \label{fig:diff_channel}
    \end{center}
\end{figure}

Figure~\ref{fig:diff_channel} separates impact for pumps organised by CPI and
by others.
For CPI-organised pumps, we see behaviour similar to Figure~\ref{fig:diff_all}: a short-term positive effect followed by a negative long-term one.
For non-CPI organised pumps there is no positive impact at all and in fact the price begins falling almost immediately relative to the market prices.

As explored in Section~\ref{subsec:dataset-analysis-and-breakdown}, CPI tend to choose coins that are more liquid, meaning they have higher volumes and are in general more visible to outside traders.
This makes pump outsiders more likely to be attracted to these CPI pumped coins since these coins that have been pumped are higher ranked and therefore more visible.
Since CPI organised pumps make up 86\% of the dataset, the prices of these have the biggest impact on the overall relative prices for the dataset, meaning that the positive short-term impact of CPI-pumps on coins is reflected in the overall average of the dataset.

Table~\ref{tab:diff_cpi} summarises the adjusted relative prices for CPI pumps
at various timeframes. The results highlight the need for multiple averages, as
the mean values for longer than 90 days are significantly different from the
median value and the IQR mean, implying an influential outlier.
From this it can be seen that coins pumped by CPI lose around 16\% of their value in the year after a pump event, relative to market prices.

\begin{table}[!htbp]
    \centering
    \caption{Summary of long term impacts of CPI organised pumps.}
    \label{tab:diff_cpi}
    \addtolength{\tabcolsep}{-0.4em}
    \begin{tabular}{lrrrrrrrr}
        \toprule
        Day & \textbf{7} & \textbf{14} & \textbf{30} & \textbf{60} & \textbf{90} & \textbf{180} & \textbf{270} & \textbf{365} \\
        \midrule
        Mean & 105.11 & 99.93 & 83.59 & 80.26 & 47.26 & 52.48 & 27.44 & 25.47 \\
        Median & 103.86 & 100.77 & 94.13 & 89.15 & 83.30 & 76.81 & 75.71 & 72.13 \\
        IQR Mean & 104.56 & 101.38 & 96.12 & 91.52 & 86.68 & 79.57 & 82.69 & 84.07 \\
        \midrule
        Average & 104.51 & 100.70 & 91.28 & 86.98 & 72.41 & 69.62 & 61.95 & 60.55 \\
        \bottomrule
    \end{tabular}
\end{table}

Table~\ref{tab:diff_others} displays the same information for non-CPI organised pumps.
The differences between the averages are much smaller than for the CPI-organised pumps, implying there are few if any big outliers in the data.
Overall coins that are pumped by non-CPI groups lose just over 25\% of their value in the year after a pump event, again relative to market prices.

\begin{table}[!htbp]
    \centering
    \caption{Summary of long term impacts of non-CPI organised pumps.}
    \label{tab:diff_others}
    \addtolength{\tabcolsep}{-0.4em}
    \begin{tabular}{lrrrrrrrr}
        \toprule
        & \textbf{7} & \textbf{14} & \textbf{30} & \textbf{60} & \textbf{90} & \textbf{180} & \textbf{270} & \textbf{365} \\
        \midrule
        Mean & 98.61 & 100.36 & 65.97 & 86.46 & 89.24 & 83.48 & 75.24 & 68.23 \\
        Median & 97.33 & 97.07 & 86.92 & 81.97 & 78.81 & 70.47 & 72.97 & 74.22 \\
        IQR Mean & 96.01 & 96.23 & 90.25 & 83.92 & 76.47 & 75.53 & 76.60 & 78.89 \\
        \midrule
        Average & 97.32 & 97.88 & 81.05 & 84.12 & 81.51 & 76.49 & 74.93 & 73.78 \\
        \bottomrule
    \end{tabular}
\end{table}

Comparing the two groups highlights that in the long-term coins pumped by both groups experience a price decrease relative to general market prices.
This makes it possible to conclude that, in general, pump events have a negative long-term impact on the value of cryptocurrencies.

\subsubsection{Impact By Number of Subsequent Pumps}
\label{subsubsec:impact-by-number-of-subsequent-pumps-cpi}

\begin{figure*}[!htpb]
    \begin{center}
        \subfloat[Mean.]{
            \includegraphics[width=0.30\linewidth]{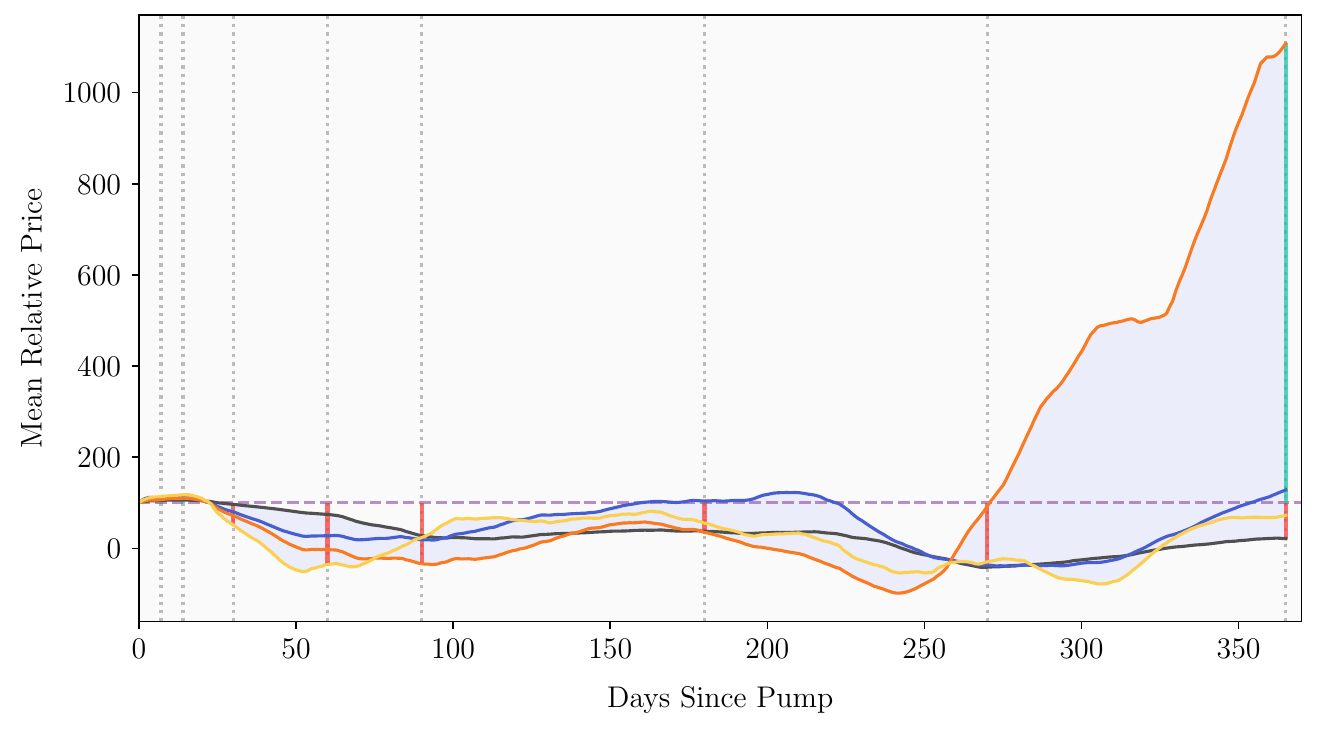}
            \label{fig:diff_cpi_full_mean}
        }
        \subfloat[Median.]{
            \includegraphics[width=0.30\linewidth]{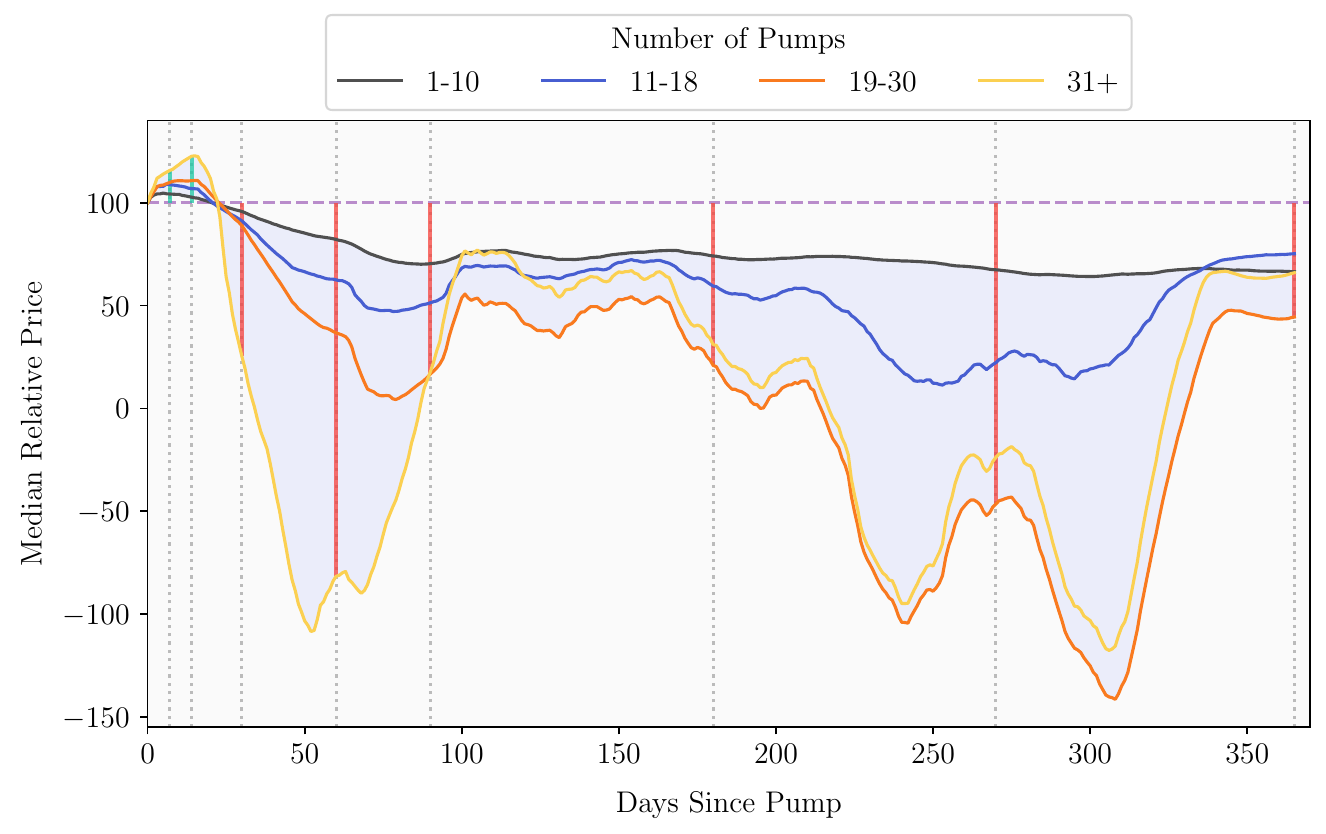}
            \label{fig:diff_cpi_full_median}
        }
        \subfloat[IQR Mean.]{
            \includegraphics[width=0.30\linewidth]{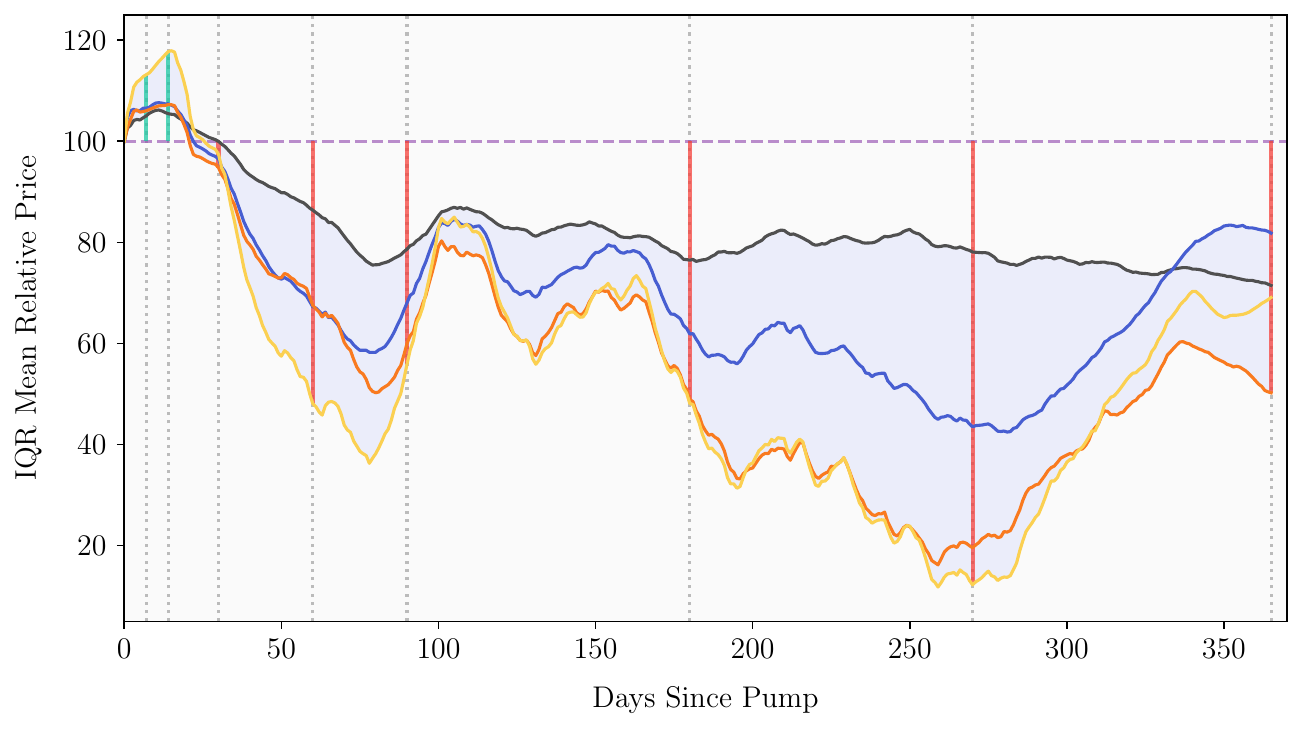}
            \label{fig:diff_cpi_full_iqr}
        }\\
        \subfloat[Mean.]{
            \includegraphics[width=0.30\linewidth]{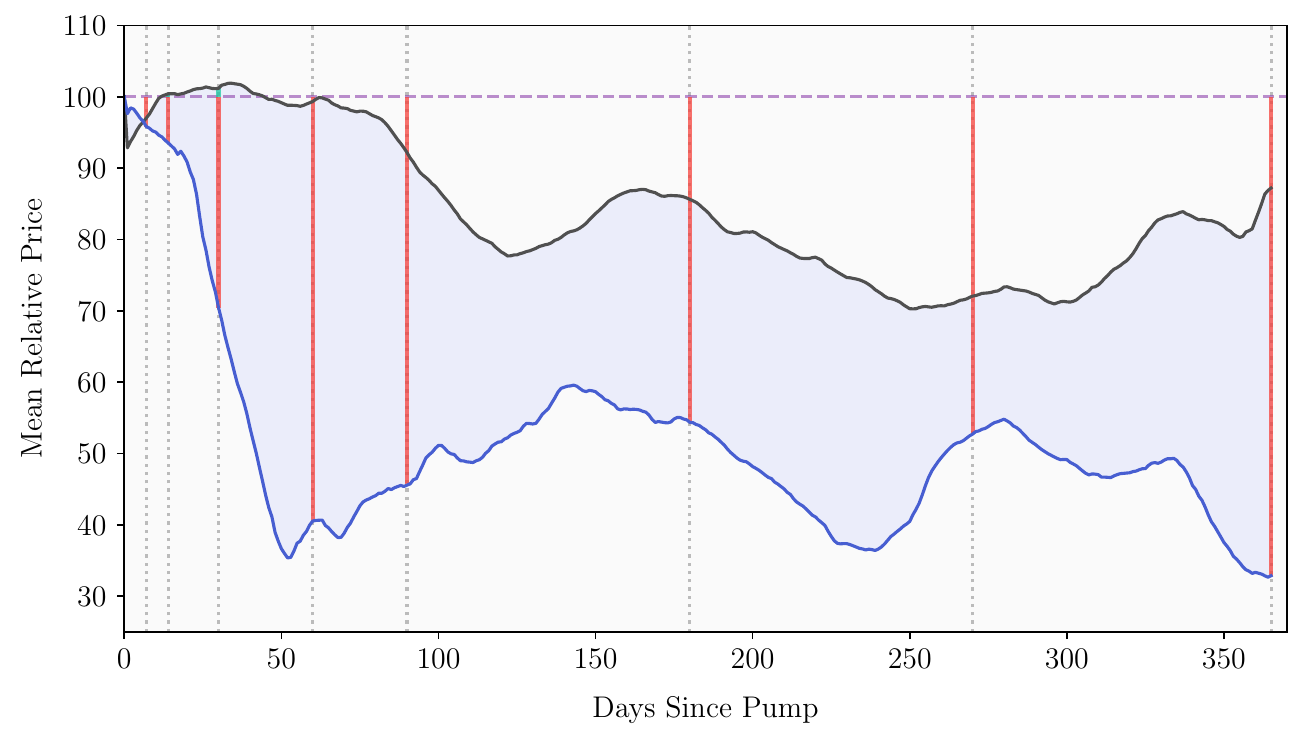}
            \label{fig:diff_others_full_mean}
        }
        \subfloat[Median.]{
            \includegraphics[width=0.30\linewidth]{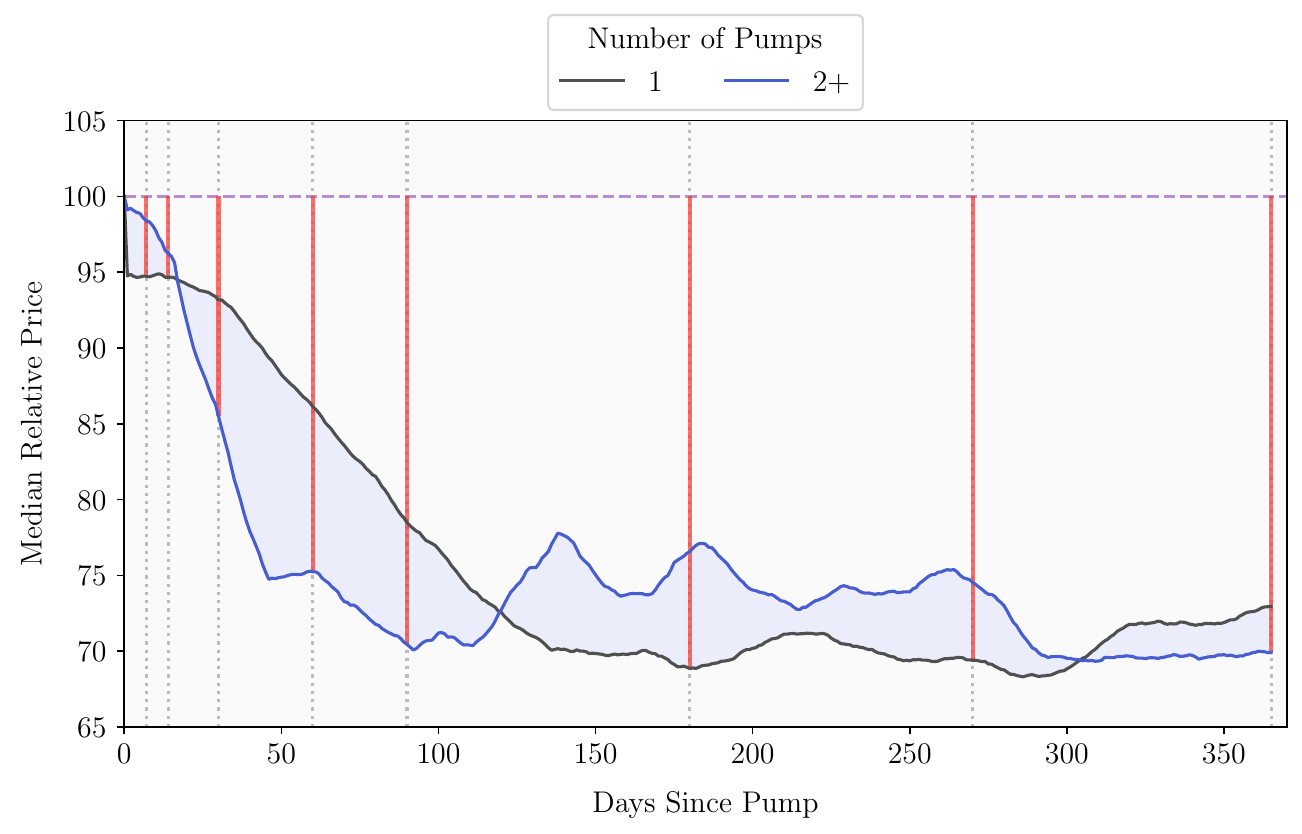}
            \label{fig:diff_others_full_median}
        }
        \subfloat[IQR Mean.]{
            \includegraphics[width=0.30\linewidth]{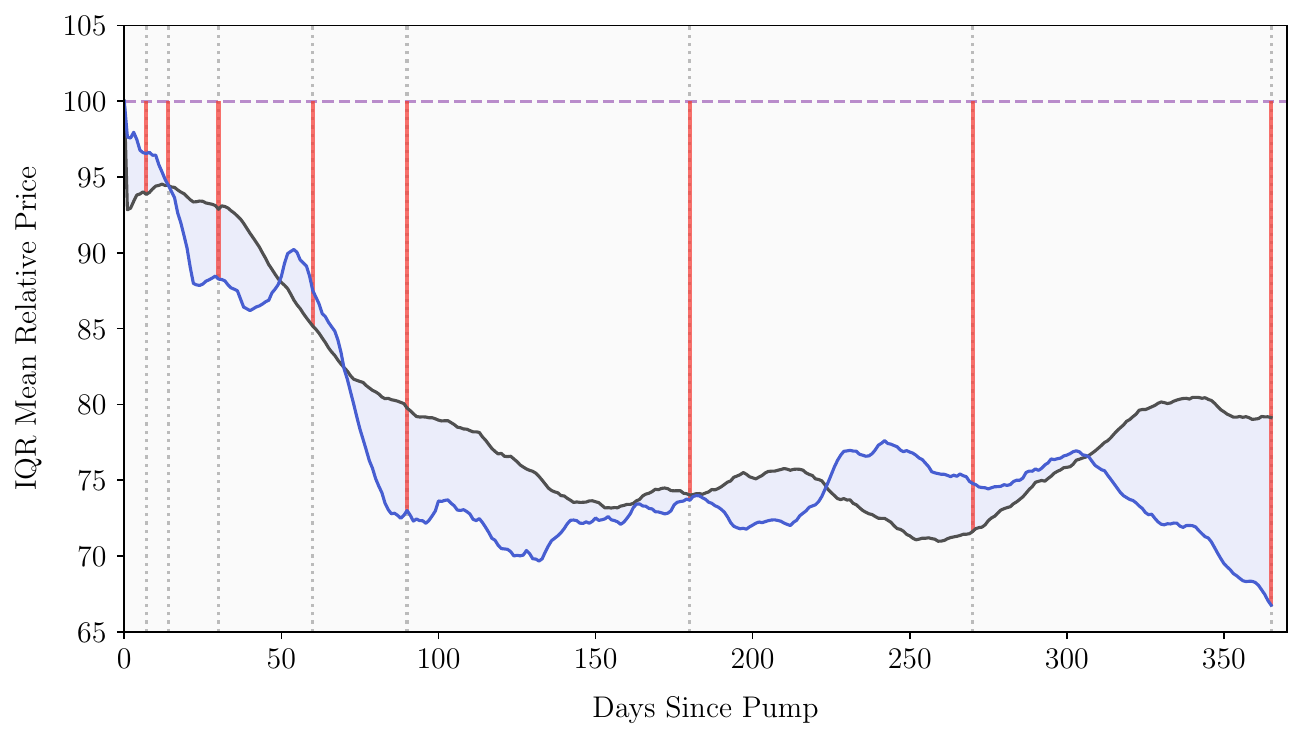}
            \label{fig:diff_others_full_iqr}
        }
        \caption{Relative prices of coins pumped by CPI (top row) and groups
other than CPI (bottom row) in the year after a pump, adjusted for general market movements and grouped by total number of pumps in the dataset.}
        \label{fig:diff_full}
    \end{center}
\end{figure*}

Figure~\ref{fig:diff_full} shows the market-adjusted long-term relative prices
of coins for the 365 days after a pump event, grouped by the total number of
pumps for unique coins in the dataset, for CPI and non-CPI organised pumps.
Looking at the mean average displayed in Figure~\ref{fig:diff_cpi_full_mean},
there appear to be outliers present with much better performance than expected, particularly in coins in the \verb|19-30| bin.
Conversely, the median average, displayed in Figure~\ref{fig:diff_cpi_full_median}, shows the potential presence of outliers that perform much worse expected.
The fact that the IQR mean (Figure~\ref{fig:diff_cpi_full_iqr}) exhibits much more similar behaviour to that found in Section~\ref{subsubsec:organiser-impact} further reinforces the notion that the other two averages are affected by outliers, at both ends of the performance spectrum.
The main pricing for non-CPI pumps also appears to be affected by outliers, since it is significantly higher for the \verb|1| pump bin and significantly lower for the \verb|2+| pump bin when compared to the other averages.
One feature that all the graphs display is the worse long-term performance of coins pumped more than once.

\begin{table}[!htbp]
    \centering
    \caption{Summary of long-term impacts of CPI and non-CPI organised pumps, grouped by number of pumps and averaged using the IQR mean.}
    \label{tab:diff_full_cpi}
    \addtolength{\tabcolsep}{-0.4em}
    \begin{tabular}{lrrrrrrrr}
        \toprule
        Day & \textbf{7} & \textbf{14} & \textbf{30} & \textbf{60} & \textbf{90} & \textbf{180} & \textbf{270} & \textbf{365} \\
        \midrule
        \multicolumn{9}{c}{CPI} \\
        \midrule
        1--10 & 105.04 & 105.47 & 99.99 & 86.38 & 78.66 & 76.47 & 78.09 & 71.47 \\
        11--18 & 106.57 & 107.42 & 96.50 & 67.27 & 68.06 & 61.85 & 43.49 & 81.83 \\
        19--30 & 106.03 & 107.27 & 94.77 & 67.53 & 59.99 & 48.87 & 19.56 & 50.23 \\
        31+ & 113.19 & 117.77 & 97.44 & 48.08 & 55.94 & 48.09 & 12.22 & 69.08 \\
        \midrule
        \multicolumn{9}{c}{non-CPI} \\
        \midrule
        1 & 93.85 & 94.45 & 92.88 & 85.16 & 79.76 & 74.02 & 71.61 & 79.13 \\
        2+ & 96.55 & 94.45 & 88.27 & 87.50 & 73.00 & 73.67 & 74.78 & 66.76 \\
        \bottomrule
    \end{tabular}
\end{table}

Table~\ref{tab:diff_full_cpi} shows a summary of the IQR mean of the
market-adjusted relative prices for coins over our selected timeframes.
This again highlights that pumps have a negative impact on the relative prices of the respective coins.
With respect to the effect of the number of pumps, between the CPI bins there is no conclusive trend, since this varies in an inconsistent manner, implying that, at least for CPI-organised pumps, the number of pumps has no easily quantifiable bearing on the price impact in this dataset.
There is divergent behaviour between the two different non-CPI bins with respect
to shorter timeframes, with the \verb|1| bin having a short-term relative price
uptick whilst the \verb|2+| bin immediately decreases.
Both categories display a relative decrease in price over the course of a year, although as discussed above the \verb|2+| has worse performance, 13\% less than the \verb|1| bin.
Whilst these features could be indicative of more pumps meaning lower long-term prices, the large range of the \verb|2+| pump bin and the lack of coins in it relative to the \verb|1| pump bin, indicate that in reality more data is needed before any conclusive interpretations can be made on this relationship.

\subsubsection{First vs the Rest}
\label{subsubsec:first-vs-rest}
Table~\ref{tab:diff_subset_cpi} shows the average adjusted price impacts of the
first and the subsequent three times a coin is pumped during CPI-organised
pumps, as outlined at the end of
Section~\ref{subsubsec:quantifying-subsequent-pumps}. Again we see the a
long-term negative impact across all the coins pumped relative to the market.
However, with regards to the variations between the groupings by pump number, a
complex picture emerges. 
The impact of additional pumps, for both CPI and non-CPI organised pumps, is hard to discern across any of the timeframes, no matter the number of pumps or re-pumps.
The only consistent trend across the pumps and re-pumps is the overall impact is negative in the long-term.


\begin{table}[!htbp]
    \centering
    \caption{Summary of long-term impacts of CPI and non-CPI organised pumps for coins on their 1st to 4th pumps, averaged using the IQR mean.}
    \label{tab:diff_subset_cpi}
    \addtolength{\tabcolsep}{-0.4em}
    \begin{tabular}{lrrrrrrrr}
        \toprule
        Days & \textbf{7} & \textbf{14} & \textbf{30} & \textbf{60} & \textbf{90} & \textbf{180} & \textbf{270} & \textbf{365} \\
        \midrule
        \multicolumn{9}{c}{CPI} \\
        \midrule
        1 & 107.80 & 109.61 & 97.55 & 69.23 & 66.15 & 58.47 & 37.29 & 68.95 \\
        2 & 107.09 & 107.35 & 99.69 & 89.27 & 71.74 & 61.41 & 50.33 & 63.10 \\
        3 & 104.87 & 104.32 & 101.66 & 97.99 & 78.42 & 71.11 & 60.08 & 63.68 \\
        4 & 104.40 & 103.30 & 101.19 & 97.36 & 81.37 & 76.51 & 66.94 & 59.19 \\
        \midrule
        \multicolumn{9}{c}{non-CPI} \\
        \midrule
        1 & 94.76 & 94.45 & 91.33 & 85.84 & 77.36 & 73.88 & 72.93 & 73.38 \\
        2 & 98.08 & 98.43 & 94.42 & 89.12 & 77.76 & 86.19 & 73.63 & 75.64 \\
        3 & 100.05 & 101.08 & 95.25 & 85.29 & 64.68 & 89.07 & 73.95 & 73.90 \\
        4 & 97.58 & 98.84 & 95.53 & 94.28 & 65.98 & 95.21 & 90.43 & 92.17 \\
        \bottomrule
    \end{tabular}
\end{table}

From this we may conclude that at least when viewed from this perspective, the
number of times a coin is pumped has no significant bearing on its long-term price.
Furthermore, this analysis in particular does not adjust for whether subsequent pumps happen in
the 365 day period after pump events, which could provide an explanation for the outlier discussed earlier.

    \section{Discussion and Conclusion}
    \label{sec:discussion-conclusion}

    \subsection{Contributions and Findings}
\label{subsec:contributions-and-findings}


We have contributed a new, enlarged dataset of cryptocurrency pump events based
on existing sources, that contains around
10,000 events, representing a 10-fold increase compared to the largest
comparable source~\cite{morgia-doge-wall-street-2021}. As well as the dataset
itself, we release the code used to collect the data, meaning the
dataset can be maintained and updated.  

Our main finding within this dataset was, relative to market prices, pump events have a long-term negative impact on the value of cryptocurrencies which represents a conclusive answer to the main hypothesis of this project.
    On average this impact is valued at a 27\% price decrease relative to market prices 365 days after a pump event.
    These findings provide specific evidence for the notion proposed by Li et al.\ that pump events are, in general, detrimental to the price of cryptocurrencies~\cite{li-pump-dump-schemes-2021}.
    Regarding the findings of Victor and Hagermann, that pump events provide a 10\% positive price impact after 100 days~\cite{victor-quantification-detection-2019}, this project finds that, after 90 days, the prices of pumped coins are over 25\% lower relative to the wider market.

Further, our investigations of the dataset reveal a \emph{quality vs quantity
tradeoff}, which exposes two different approaches to organising pumps.
    The quality approach involves choosing coins with lower liquidity, and therefore lower value, for pumps at lower frequencies which achieve a high percentage price increase.
    The quantity approach, on the other hand, focuses on coins with higher liquidity, and therefore higher value, for pumps at higher frequencies which achieve lower percentage price increases.
    Further empirical evidence, in the form of price and volume separation, was found through analysing the market data for each grouping.

A significant event within our data was the collapse of the cryptocurrency
exchange FTX. An additional analysis presented in Appendix~\ref{subsec:the-collapse-of-ftx} covers the
impact of this event. We find the collapse had no material direct impact on the
number of pumps organised. However, the collapse impacted the underlying prices 
and volumes of cryptocurrencies in general, this in turn impacted the value of
coins traded during pump events, decreasing the amount of money that operators,
and to a lesser extent participants, could potentially make. 

These findings highlight the need for further regulation of pump and dumps in the cryptocurrency sphere, given that these schemes are in effect market manipulation and are found to have had a significant negative impact on the value of cryptocurrencies.
Regulators and governments are beginning to accept that pump and dump schemes are a significant problem that should be addressed in any future cryptocurrency regulation; a recent UK government consultation cited them as a target for potential market abuse regulation~\cite{UK-gov-crypto-def-2023}.
However, this lack of regulation with respect to pump and dump schemes, which was first highlighted as a regulatory issue in 2018~\cite{hamrick-pump-dump-economics-2018}, coupled with the high frequency and relative anonymity of such events mean they will continue to pervade cryptocurrency markets until decisive action is taken.

\subsection{Future Work}
\label{subsec:future-work}
There is significant scope for future work, some suggestions for which are discussed below.

\paragraph{Investigating more channels}
Only the 130 most suspicious out of 800 channels were investigated from the initial list from PumpOlymp.
Whilst these channels still resulted in a dataset far larger than existing ones,
there are still a significant number of channels which may fruitfully be
monitored for more pumps. Future work could expand the dataset by investigating
these sources.

\paragraph{Pumps on DEX}
As the APIs used to retrieve the market OHLCV data are for centralised exchanges, any pumps found taking place on decentralised exchanges could not be analysed and were therefore ignored.
Additional work could focus on probing pump events on DEXs by analysing on-chain exchange transactions, using a method such as that described by Li et al.\ \cite{li-pump-dump-schemes-2021}.

\paragraph{Yobit and further data}
As discussed in Section~\ref{subsubsec:aggregating-pump-events-into-a-single-table}, pumps organised on Yobit could not be analysed due to its API only providing 7 days of OHLCV data, which excluded around 350 pumps from further analysis.
Subsequent research could investigate these by using data from source such as Kaiko, which provides over 10 years of historical data for over 100 exchanges including Yobit~\cite{kaiko-about-2023}\footnote{The reason Kaiko has Yobit data is that it collects pricing data for all cryptocurrencies listed on an exchange in real time.}.
Furthermore, a limitation of the methodology used for this project was the OHLCV data was not checked for missing dates, due to time constraints, which again could be solved through the use of Kaiko data.

\paragraph{General market data calculations}
The method used to calculate the relative market prices relied on the top 10 cryptocurrencies by market capitalisations as of 22nd April 2023.
Whilst this provided a way of easily measuring the performance of pumped coins relative to the wider cryptocurrency market, it does not take into account previous changes to this top 10 ranking\footnote{For example, DOGE only achieved a top 10 market captitalisation in 2021.}, meaning they may not represent the general market movements for earlier pump events.
A solution to this would be to use historic CoinMarketCap market capitalisation to retrieve the top 10 coins at the time of each pump event and then use the data for these in order to calculate adjusted relative prices.



\subsection{Conclusion}
\label{subsec:conclusion}
In summary, our work has produced a new and extended dataset of pump events and provided a detailed breakdown and analysis of the performance of these pump events.
Exploring the long-term impacts of pump events on cryptocurrencies reveals an overall negative impact over a 365 day period with respect to general market prices.
The analysis also highlighted different tactics in the form of a \emph{quality
vs quantity} tradeoff, with different targets and different forms of market
impact.

    \appendices
    \section{The Collapse of FTX}
\label{subsec:the-collapse-of-ftx}
The collapse of FTX had a material impact on both pump and dump operations and cryptocurrencies markets in general.
Whilst this collapse is too recent\footnote{FTX filed for bankruptcy on the 11th November 2022.} to apply the long-term analysis used in Section~\ref{subsec:long-term-impacts}, it offers an opportunity to explore the direct impact of this major event on pump events.

\subsubsection{Number of Pump Events}
\label{subsubsec:number-of-pumps-events-ftx}
\begin{figure}[!htbp]
    \centering
    \includegraphics[width=0.45\textwidth]{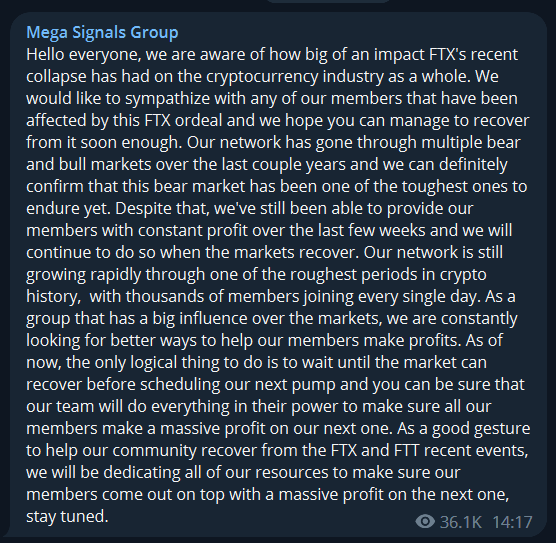}
    \caption{An attempt by a pump organiser to reassure participants after the collapse of FTX.}
    \label{fig:ftx-impact-message}
\end{figure}

The first impact explored is the number of pump events organised.
As evidenced by Figure~\ref{fig:ftx-impact-message}, the collapse of FTX caused some channel admins to pause the organisation of pumps whilst the market ``recovered''\footnote{Quote taken from Figure~\ref{fig:ftx-impact-message}.}.

Table~\ref{tab:proportion_ftx} shows the proportion of pumps that happened before and after the collapse of FTX as well as the proportion of time each period represents.
These proportions indicate that the spread of pumps before and after the collapse of FTX is roughly the same, implying that the collapse had little impact on the number of pumps.

\begin{table}[!htbp]
    \centering
    \caption{Proportion of pumps before and after the collapse of FTX.}
    \label{tab:proportion_ftx}
    \begin{tabular}{lrrrr}
        \toprule
        {} & \textbf{Number} & \textbf{Proportion} & \textbf{ of Time} & \textbf{Pumps per Week} \\
        \midrule
        Before & 8725 & 94.93 & 94.32 & 33.12 \\
        After  &  466 &  5.07 &  5.68 & 29.39 \\
        \bottomrule
    \end{tabular}
\end{table}

\begin{table}[!htbp]
    \centering
    \caption{Pumps separated by exchange before and after the collapse of FTX.}
    \label{tab:proportion_exchange_ftx}
    \begin{tabular}{lrrrrr}
        \toprule
        {} & \textbf{Binance} & \textbf{Hotbit} & \textbf{Kucoin} \\
        \midrule
        Before &    96.45 &   72.05 &   77.94 \\
        After  &     3.55 &   27.95 &   22.06 \\
        \bottomrule
    \end{tabular}
\end{table}

Separating out by exchanges, shown in Figure~\ref{tab:proportion_exchange_ftx}, reveals that the proportion of pumps organised on Binance, and therefore those organised by CPI\footnote{Since this is the dominant channel organising pumps on Binance.}, decreased relative to the time periods before and after.
Bitmart and Bittrex have been excluded as there are no pumps organised on these exchanges post-collapse.
In reality, this means that there were fewer pumps performed per unit of time after the collapse of FTX than before, indicating a negative impact on the number of pumps on Binance after the collapse.
This behaviour perhaps reflects the scrutiny directed towards Binance after the collapse of FTX, particularly when the auditor Mazars stopped providing proof of reserve checks in December 2022~\cite{guardian-binance-2022}, which led to over \$6 billion of customer deposits being withdrawn in a week, severly affecting the confidence of both the market and users in the exchange.

Conversely, there were more pumps performed on Hotbit and Kucoin per unit of time after the collapse, suggesting that there was a positive impact on the number of pumps.
Since cryptocurrency markets in general have recovered in 2023, for example the price of BTC is up over $80\%$\footnote{As of the 12th April 2023.} compared to the start of 2023~\cite{cmc-bitcoin-prices-2023}, which indicates investor confidence in cryptocurrencies in general is returning.
In turn, this means that there are more potential participants for organisers of pump and dump schemes to recruit, making it attractive for them to run more pumps, causing the increase.

\subsubsection{Short Term Impacts}
\label{subsubsec:ftx-short-term-impacts}
\begin{figure}[!htbp]
    \centering
    \includegraphics[width=0.5\textwidth]{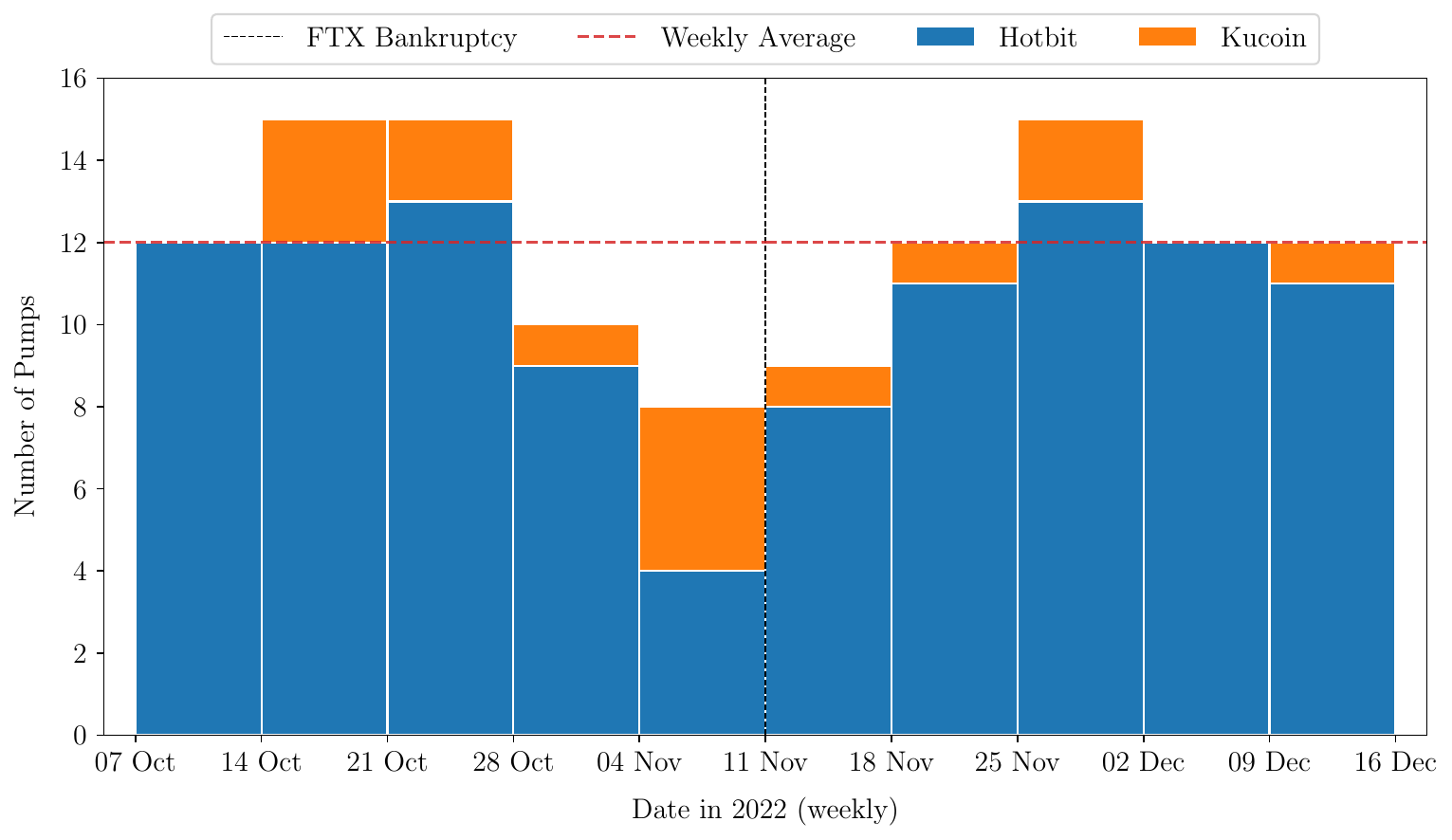}
    \caption{Number of pumps around the collapse of FTX on Hotbit and Kucoin.}
    \label{fig:hotbit_kucoin_ftx_close}
\end{figure}

As discussed above, there are other forces at play other than the collapse of FTX that have been affecting pump and dumps, particularly in the medium term since its collapse.
Exploring the 5 weeks either side of the collapse offers a more direct and focused look at the immediate impacts.
Figure~\ref{fig:hotbit_kucoin_ftx_close} shows the number of pumps on Hotbit and Kucoin before and after the collapse of FTX, with the average number of pumps in the periods before and after indicated by the red lines.
This line indicates that this average was the same before and after the collapse, but there is a clear drop in the immediate weeks either side of it.
The impact of uncertainty surrounding FTX emerged around a week before it declared bankruptcy, explaining the drop in the week preceding its collapse, and the collapse itself explains the drop in the week after.

The recovery post-collapse was swift, with the number of pumps returning to the average number only a week after the collapse, particularly driven by pumps on Hotbit.
Hotbit is a much smaller exchange than FTX was and operates in somewhat suspicious fashion\footnote{For example the entire exchange had its assets frozen for over a month from August 2022 during an investigation by law enforcement, meaning customers could not deposit, withdraw or trade funds~\cite{hotbit-suspend-2022, hotbit-unsuspend-2022} and it has virtually no KYC requirements~\cite{hotbit-kyc-2023}.} so participants in pumps organised on Hotbit had likely already experienced uncertainty around access to their assets.
It could also be caused by the specific pump operators not caring about the collapse and trying to attract new participants with an alternative way to make money from cryptocurrencies.

\begin{figure}[!htbp]
    \centering
    \includegraphics[width=0.5\textwidth]{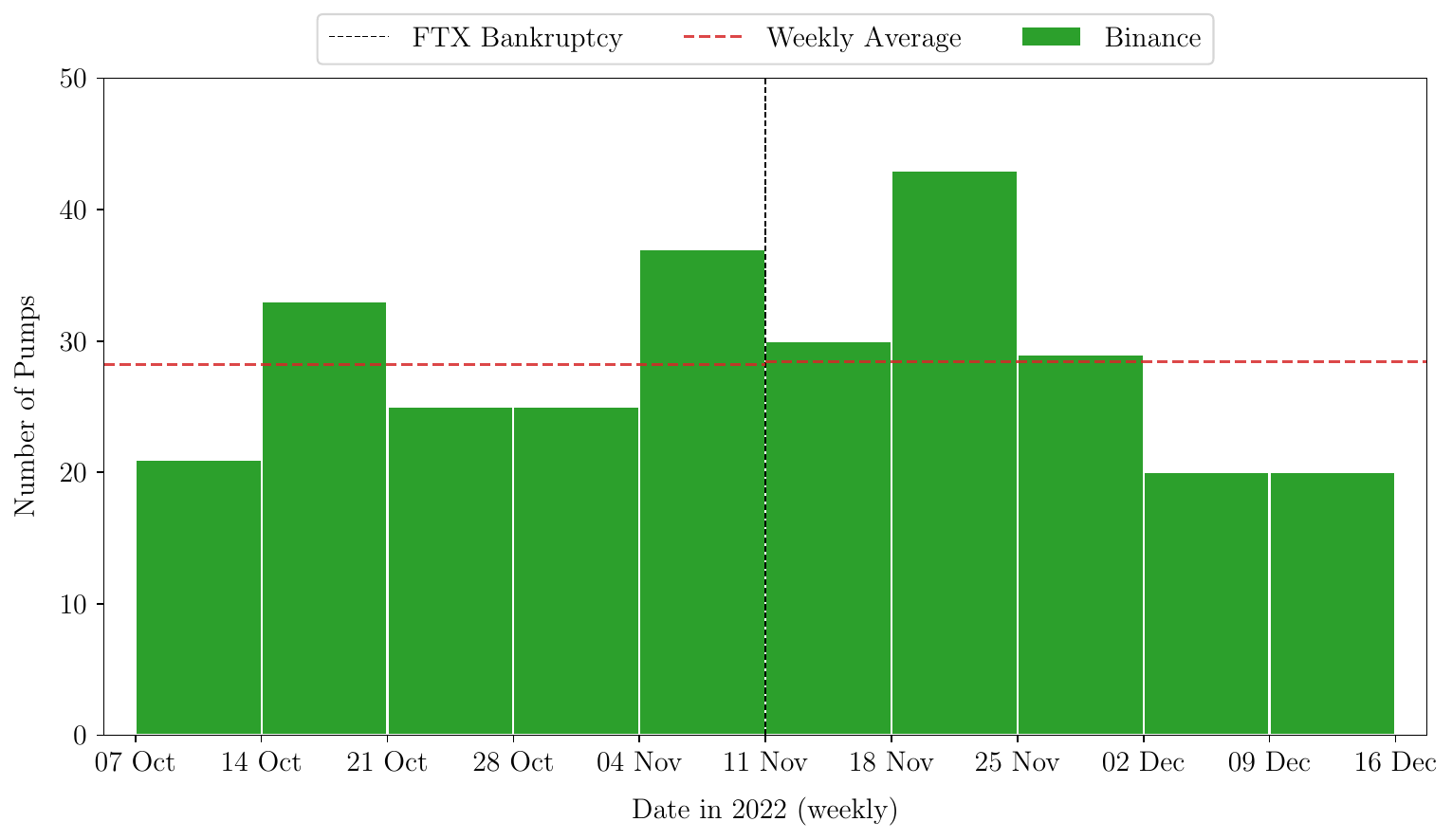}
    \caption{Number of pumps around the collapse of FTX on Binance.}
    \label{fig:binance_ftx_close}
\end{figure}

In contrast Figure~\ref{fig:binance_ftx_close} shows the number of pumps on Binance around the collapse.
This indicates that the number of pumps was higher in the week immediately prior to the collapse compared to those preceding it.
Since all these pumps were organised by CPI, who target lower returns from a higher quantity of pumps, this could be potentially an attempt by the pump organisers to capitalise on the market uncertainty.
The biggest drop for the Binance organised pumps comes at the start of December, when there was $\$6$ billion of withdrawals in a week due to the uncertainty around its proof of reserves report~\cite{reuters-binance-2022}, since this directly affected the platform that participants were using.

Overall the collapse of FTX appeared to have varying levels of impact on the amount of pump and dump schemes organised around the collapse, from a short-term increase for CPI organised pumps and a very short-term drop for the others.
The fact that none of the pumps were organised on FTX would have also meant that they were only affected indirectly by the fallout, potentially limiting the impact of the collapse.

\subsubsection{Performance of Pump Events}
\label{subsubsec:performance-of-pump-events-ftx}
Moving on to the performance of pump events before and after the collapse of FTX, Table~\ref{tab:ftx_performance_impact} shows the difference between the maximum price increase and volume moved during pump events before and after the collapse.
As established in Section~\ref{subsec:pump-performance}, there are big differences between these values for CPI and non-CPI organised pumps, so the data is further separated by this.

\begin{table}[!htbp]
    \centering
    \caption{Pump performance before and after the collapse of FTX.}
    \label{tab:ftx_performance_impact}
    \begin{tabular}{lrrrr}
        \toprule
        {} & \multicolumn{2}{c}{\textbf{Price Increase}} & \multicolumn{2}{c}{\textbf{Relative Pump Volume}} \\
        {} & {CPI} & {Non-CPI} & {CPI} & {Non-CPI} \\
        \midrule
        Before & 6.72 & 434.22 & -95.95 & 4284.40 \\
        After & 25.92 & 688.16 & -77.49 & 2006.64 \\
        \bottomrule
    \end{tabular}
\end{table}
  -

Both the CPI and non-CPI organised pumps have higher average price increases post the collapse of FTX, however they differ on the proportion of the pre-pump volume and the volume moved during the pump represents.

Across 2022 the amount of trading volume across crypto exchanges reduced, particularly towards the end of year~\cite{crytponews-binance-2022}.
This would have significantly affected CPI organised pumps since they tend to target coins with higher liquidity, i.e.\ coins with more trading volume by non-participating market actors, which are likely to be those that have reduced their trading volume.
These lower trading volumes by non-participating market actors causes the pre-pump volume to be on average lower hence meaning that for a given pump volume the proportion of the pre-pump volume this represents increases.
This is further shown in Table~\ref{tab:ftx_pre_pump} which displays that there is a much bigger fall in the pre-pump volume than in the during pump volume for CPI-organised pumps, meaning that the relative proportion of the during pump volume increases compared to the pre-pump volume, causing the behaviour in Table~\ref{tab:ftx_performance_impact}.

\begin{table}[!htbp]
    \centering
    \caption{Relative difference of pre and during pump prices and volumes before and after the collapse.}
    \label{tab:ftx_pre_pump}
    \begin{tabular}{lrrrr}
        \toprule
        {} & \multicolumn{2}{c}{\textbf{Pre-Pump}} & \multicolumn{2}{c}{\textbf{During Pump}} \\
        {} & {CPI} & {Non-CPI} & {CPI} & {Non-CPI} \\
        \midrule
        Price  & -43.65 & -71.77 & -33.51 & -58.34 \\
        Volume & -86.02 & -26.62 & -22.39 & -64.74 \\
        \bottomrule
    \end{tabular}
\end{table}

For non-CPI organised pumps the average pump volume fell by a significant amount, particularly relative to the pre-pump falls.
Since this decrease in the during pump volume is greater than decrease in the pre-pump one, and this represents the decrease due to general market factors, this implies fewer participants or at least less volume per participant as this figure is far lower than the pre-pump figure.

Looking back at the relative price increases in Table~\ref{tab:ftx_performance_impact} it appears that there is a higher relative maximum price increase post-collapse, which implies that pumps performed better.
However, Table~\ref{tab:ftx_pre_pump} shows the reason for this is the maximum price of a pump fell less than the pre-price, which gives a bigger percentage return at a lower value.
Furthermore it is easier to achieve a higher percentage increase on a lower price coin, further suggesting that much of this perceived increase in percentage returns is not valid.

\begin{figure}[!htbp]
    \centering
    \includegraphics[width=0.5\textwidth]{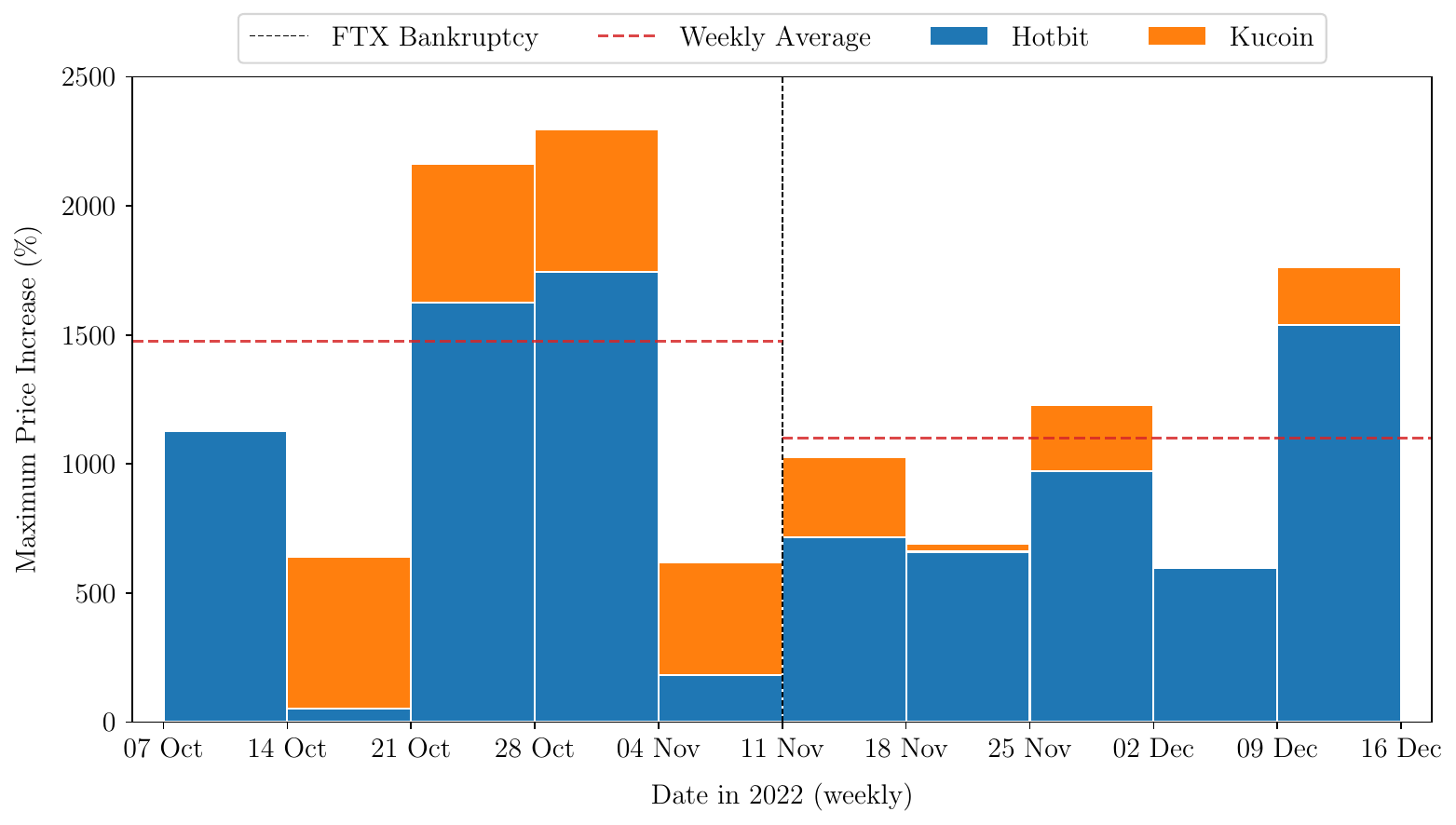}
    \caption{Maximum price increase of pumps around the collapse of FTX on Hotbit and Kucoin.}
    \label{fig:hotbit_kucoin_price_close}
\end{figure}

Figure~\ref{fig:hotbit_kucoin_price_close} displays the percentage price increases for pumps on Hobit and Kucoin for the weeks immediately surrounding the collapse of FTX\@.
This shows that the average percentage price increase decreased after the collapse, although there is significant variation in the weeks prior to it, suggesting this could be driven by factors other than just the collapse.

\begin{figure}[!htbp]
    \centering
    \includegraphics[width=0.5\textwidth]{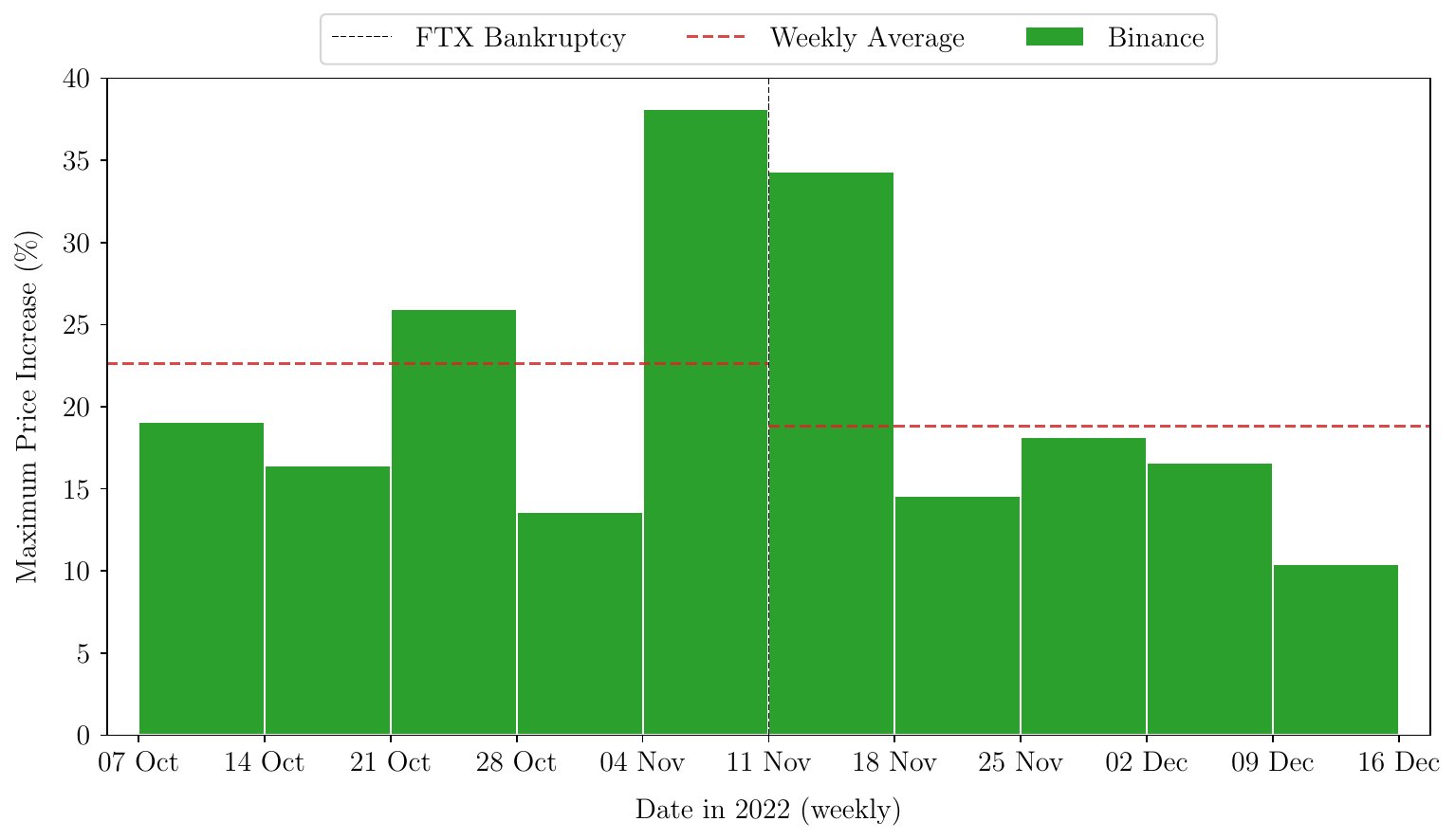}
    \caption{Maximum price increase of pumps around the collapse of FTX on Binance.}
    \label{fig:binance_price_close}
\end{figure}

Figure~\ref{fig:binance_price_close} shows the same again for pumps on Binance.
Whilst the average is again lower post-collapse, there is a significant uptick in performance in the week immediately prior to and after the collapse, potentially caused by investors moving volume from FTX to Binance as its troubles became more apparent and investors moved their funds to other exchanges.
Since Binance is by far the biggest exchange by volume~\cite{cmc-exchanges-2023}, even before the collapse of FTX, it would be the main destination for such moves and as the coins chosen by CPI (and therefore the ones pumped on Binance), are more liquid, they would be more likely to receive volume from such trades.

\subsubsection{Total Value}
\label{subsubsec:total-value}
As established above, there were changes in both the price and volume of pump events after the collapse of FTX\@.
The best way to establish the monetary impact of these effects is to look at the changes in value caused by pump events\footnote{As established in Section~\ref{subsubsec:value-added}.}.
Table~\ref{tab:ftx_overall_value} shows the value of units moved in the pre-pump and during pump periods.

\begin{table}[!htbp]
    \centering
    \caption{Total value of cryptocurrency moved }
    \label{tab:ftx_overall_value}
    \begin{tabular}{lrrrr}
        \toprule
        {} & \multicolumn{2}{c}{\textbf{CPI}} & \multicolumn{2}{c}{\textbf{Non-CPI}} \\
        {} & Pre-Pump & During Pump & Pre-Pump & During Pump \\
        \midrule
        Before & \num{2.62e9} & \num{1.13e8} & \num{1.56e8} & \num{3.66e10} \\
        After  & \num{2.07e8} & \num{5.84e7} & \num{3.24e7} &  \num{5.38e8} \\
        \midrule
        \textbf{Change} & -79.12 & -85.31 & -92.12 & -48.40 \\
        \bottomrule
    \end{tabular}
\end{table}

Overall whilst the value of units moved during pumps shows some resilience with respect to the change in the general value of units moved in the pre-pump periods, it still represents a significant reduction compared to the pre-collapse values.
Operators, and more broadly some participants, make money off these schemes based on the change in value of the coin.
Hence, these decreases represent  a large reduction in the amount of monetary value that can be made off these schemes, in turn reducing their performance in this respect.

\vfill\eject
    \section{Data Sources}
\label{appx:data-sources}

\begin{table}[!htbp]
    \centering
    \caption{List of exchanges used and their respective pump count and links.}
    \label{tab:exchanges}
    \begin{tabular}{lll}
        \toprule
        \textbf{Exchange} & \textbf{Pumps} & \textbf{Link} \\
        \midrule
        Binance &   9129 &    \url{https://www.binance.com} \\
        BitMart &     15 &    \url{https://www.bitmart.com} \\
        Bittrex &     25 & \url{https://global.bittrex.com} \\
        Hotbit  &    852 &      \url{https://www.hotbit.io} \\
        Kucoin  &     92 &     \url{https://www.kucoin.com} \\
        YoBit   &    351 &       \url{https://yobit.net/en} \\
        \bottomrule
    \end{tabular}
\end{table}

\begin{table}[!htbp]
    \centering
    \caption{List of Telegram channels used and their respective channel codes, number of pump events and links.}
    \label{tab:telegram_channels}
    \addtolength{\tabcolsep}{-0.4em}
    \scalebox{0.9}{
        \begin{tabularx}{0.55\textwidth}{@{}lXX@{}}
            \toprule
            \textbf{Channel Code} & \textbf{Group Name} & \textbf{Telegram Link} \\
            \midrule
            ATW &                       Alt the Way &                   \url{https://t.me/AltTheWay} \\
            BCP &                   Big Crypto Pump &       \url{https://t.me/bigcryptocurrencypump} \\
            BPG &                    Big Pump Group &            \url{https://t.me/bigpumpgroup_com} \\
            PBS &                 Big Pump Shitcoin &            \url{https://t.me/Bigpump_shitcoin} \\
            BPS &                   Big Pump Signal &               \url{https://t.me/bigpumpsignal} \\
            BPC &              Biggest Pump Channel &    \url{https://t.me/Biggestcryptopumpchannel} \\
            BP2 &                      Binance 24/7 &                  \url{https://t.me/binance247} \\
            BCS &            Binance Crypto Signals &          \url{https://t.me/CryptoSignals_pump} \\
            BPF &               Binance Pump Family &            \url{https://t.me/rocketpumptrader} \\
            BPD &             Binance Pump and Dump &       \url{https://t.me/binance_pump_and_dump} \\
            BGW & Binance/GateIO/Kucoin Whales Pump &           \url{https://t.me/binacepumpswhales} \\
            CCS &                Coin Coach Signals &            \url{https://t.me/CoinCoachSignals} \\
            CP &                     Cryptic Pumps &                \url{https://t.me/CrypticPumps} \\
            CCB &                     Crypto Coin B &            \url{https://t.me/CryptoCoinsCoach} \\
            CCPS &          Crypto Coin Pump Signals &       \url{https://t.me/Cryptocoinpumpsignals} \\
            CP3 &                   Crypto Pump 360 &               \url{https://t.me/cryptopump360} \\
            CPC &                  Crypto Pump Club &              \url{https://t.me/cryptoclubpump} \\
                CPI &                Crypto Pump Island &          \url{https://t.me/crypto_pump_island} \\
            CSP &               Crypto Signal Pumps &               \url{https://t.me/crypto_pump07} \\
            CW &                      Crypto Waves &            \url{https://t.me/CryptoCoinsWaves} \\
            C4P &                      Crypto4Pumps &                \url{https://t.me/Crypto4Pumps} \\
            CPS &               Crytopia Pump Squad &          \url{https://t.me/cryptoflashsignals} \\
            FCS &               Fast Crypto Signals &                   \url{https://t.me/fastcrypt} \\
            HPA &                   Hit Pump Angels &               \url{https://t.me/hitpumpangels} \\
            HEP &               Hotbit English Pump &         \url{https://t.me/hotbit_english_pump} \\
            HPF &                  Hotbit Pump Free &              \url{https://t.me/hotbitpumpfree} \\
            HPS &               Hotbit Pump Signals &          \url{https://t.me/hotbit_pump_signal} \\
            HTP &               Hotbit Trading Pump &         \url{https://t.me/binancepumpchannell} \\
            KCPT &       Kucoin Crypto Pumps Trading & \url{https://t.me/kucoin_crypto_pumps_trading} \\
            KPW &                   Kucoin Pump WSB &              \url{https://t.me/kucoinpumpswsb} \\
            LUX &                         Luxurious &             \url{https://t.me/LuxuriousCrypto} \\
            MPC &               Magic Pumps Channel &                  \url{https://t.me/magic_pump} \\
            MCP &                  Mega Crypto Pump &              \url{https://t.me/pump_and_dumpp} \\
            MPG &                   Mega Pump Group &             \url{https://t.me/mega_pump_group} \\
            PKG &               Pump King Community &                 \url{https://t.me/pumpingking} \\
            PL &                        Pump Leaks &                   \url{https://t.me/pumpleaks} \\
            RPB &              Rocket Pumps Binance & \url{https://t.me/exclusivepumpanddumpbinance} \\
            SP &                       Softex Pump &                  \url{https://t.me/softexpump} \\
            TWP &                     Today We Push &                 \url{https://t.me/TodayWePush} \\
            TCC &              Trading Crypto Coach &          \url{https://t.me/tradingcryptocoach} \\
            TCG &              Trading Crypto Guide &                   \url{https://t.me/TCGFORYOU} \\
            WSBP &               WallStreetBet Pumps &             \url{https://t.me/wallstbetspumps} \\
            WCG &               Whales Crypto Guide &                 \url{https://t.me/Whalesguide} \\
            YP &                       Yobit Pumps &                \url{https://t.me/yobitpump_en} \\
            \bottomrule
        \end{tabularx}
    }
\end{table}

\end{document}